\documentclass[12pt,letterpaper]{article}
\usepackage{amsmath,amssymb,array,calc,rotating,epsfig,psfrag,amscd, cite}

\makeatletter
\renewcommand\section{\@startsection {section}{1}{\z@}%
                                   {-3.5ex \@plus -1ex \@minus -.2ex}
                                   {2.3ex \@plus.2ex}%
                                   {\normalfont\large\bfseries}}
\renewcommand\subsection{\@startsection{subsection}{2}{\z@}%
                                     {-3.25ex\@plus -1ex \@minus -.2ex}%
                                     {1.5ex \@plus .2ex}%
                                     {\normalfont\bfseries}}
\makeatother

\let\non\nonumber

\let\s=\sigma

\newcommand{\bea}{\begin{eqnarray}}
\newcommand{\eea}{\end{eqnarray}}
\newcommand{\be}{\begin{equation}}
\newcommand{\ee}{\end{equation}}

\newcommand{\p}{\partial}

\newcommand{\C}[1]{$(\ref{#1})$}

\def\IZ{\relax\ifmmode\mathchoice
{\hbox{\cmss Z\kern-.4em Z}}{\hbox{\cmss Z\kern-.4em Z}}
{\lower.9pt\hbox{\cmsss Z\kern-.4em Z}} {\lower1.2pt\hbox{\cmsss
Z\kern-.4em Z}}\else{\cmss Z\kern-.4em Z}\fi}
\def\IR{\relax{\rm I\kern-.18em R}}

\def\one{{\hbox{ 1\kern-.8mm l}}}

\newlength{\bredde}
\def\slash#1{\settowidth{\bredde}{$#1$}\ifmmode\,\raisebox{.15ex}{/}
\hspace*{-\bredde} #1\else$\,\raisebox{.15ex}{/}\hspace*{-\bredde}
#1$\fi}

\newsavebox{\zzzbar}
\sbox{\zzzbar}
  {\setlength{\unitlength}{0.9em}
  \begin{picture}(0.6,0.7)
  \thinlines
  \put(0,0){\line(1,0){0.6}}
  \put(0,0.75){\line(1,0){0.575}}
  \multiput(0,0)(0.0125,0.025){30}{\rule{0.3pt}{0.3pt}}
  \multiput(0.2,0)(0.0125,0.025){30}{\rule{0.3pt}{0.3pt}}
  \put(0,0.75){\line(0,-1){0.15}}
  \put(0.015,0.75){\line(0,-1){0.1}}
  \put(0.03,0.75){\line(0,-1){0.075}}
  \put(0.045,0.75){\line(0,-1){0.05}}
  \put(0.05,0.75){\line(0,-1){0.025}}
  \put(0.6,0){\line(0,1){0.15}}
  \put(0.585,0){\line(0,1){0.1}}
  \put(0.57,0){\line(0,1){0.075}}
  \put(0.555,0){\line(0,1){0.05}}
  \put(0.55,0){\line(0,1){0.025}}
  \end{picture}}

\newcommand{\ena}{\end{eqnarray}}
\newcommand{\beqa}{\begin{eqnarray}}
\newcommand{\eeqa}{\end{eqnarray}}

\def\s{\sigma}

\begin{document}
\begin{titlepage}

\begin{center}



\vskip 2 cm
{\Large \bf Perturbative type II amplitudes for BPS interactions}\\
\vskip 1.25 cm { Anirban Basu\footnote{email address:
    anirbanbasu@hri.res.in} } \\
{\vskip 0.5cm Harish--Chandra Research Institute, Chhatnag Road, Jhusi,\\
Allahabad 211019, India\\}

\end{center}

\vskip 2 cm

\begin{abstract}
\baselineskip=18pt

We consider the perturbative contributions to the $\mathcal{R}^4$, $D^4\mathcal{R}^4$ and $D^6\mathcal{R}^4$ interactions in toroidally compactified type II string theory. These BPS interactions do not receive perturbative contributions beyond genus three. We derive Poisson equations satisfied by these moduli dependent string amplitudes. These T--duality invariant equations have eigenvalues that are completely determined by the structure of the integrands of the multi--loop amplitudes. The source terms are given by boundary terms of the moduli space of Riemann surfaces corresponding to both separating and non--separating nodes. These are determined directly from the string amplitudes, as well as from U--duality constraints and logarithmic divergences of maximal supergravity. We explicitly solve these Poisson equations in nine and eight dimensions.

\end{abstract}

\end{titlepage}


\section{Introduction}

The effective action of string theory in a fixed background contains invaluable information about the S--matrices of the theory in that classical background (for theories with self--dual field strengths we mean covariant equations of motion). This includes information about the various perturbative as well as non-perturbative contributions to the S--matrix elements. This elucidates the role of the non--perturbative U--duality symmetry of the theory. Such analysis can be done in considerable detail for BPS interactions in toroidal compactifications of type II string theory which preserve maximal supersymmetry. These are the $\mathcal{R}^4$, $D^4\mathcal{R}^4$ and $D^6\mathcal{R}^4$ interactions, which are $1/2$, $1/4$ and $1/8$ BPS interactions respectively in the low energy effective action~\cite{Green:1997tv,Green:1997as,Kiritsis:1997em,Green:1998by,Green:1999pu,Green:2005ba,Berkovits:2006vc,Basu:2007ru,Basu:2007ck,Green:2008bf,Basu:2008cf,Green:2010kv,Green:2010wi,Basu:2011he,D'Hoker:2013eea,Gomez:2013sla,D'Hoker:2014gfa,Basu:2014hsa,Bossard:2014aea,Pioline:2015yea,Bossard:2015uga,Bossard:2015oxa}. Consequently this generalizes to interactions lying in the same supermultiplets as these ones. The coefficient functions of these various four graviton interactions which depend on the string coupling as well as the moduli of the compactification are modular forms that are invariant under U--duality in the Einstein frame. When expanded around weak string coupling, they yield various perturbative contributions to the S--matrix. In the string frame, the perturbative contribution at genus $g$ has an overall factor of the form $(e^{-2\phi} V_d)^{1-g}$, where $\phi$ is the ten dimensional dilaton, and $V_d$ is the volume of the target space torus $T^d$ in the string frame metric. The remaining moduli dependence of the perturbative S--matrix element is encoded in a T--duality invariant coefficient which depends only on $G_{ij}$ and $B_{ij}$, which are the components of the metric and the NS--NS two form along $T^d$. Hence calculating these coefficients is of importance not only for analyzing the perturbative S--matrices of these interactions, but also to understand the role of U--duality.       

While generic interactions receive perturbative contributions to all orders, the BPS interactions satisfy certain non--renormalization theorems. The $\mathcal{R}^4$ interaction receives perturbative contributions only upto genus one. The $D^4\mathcal{R}^4$ and $D^6\mathcal{R}^4$ interactions receive such contributions upto genus two and three respectively. The perturbative contributions to these interactions form the central theme of our analysis. They are given by an integral over the supermoduli space of super--Riemann surfaces, which for the cases we consider, reduce to integrals over the moduli space of Riemann surfaces. Such integrals that arise in string theory have been generally studied in various cases, and sometimes they can be explicitly evaluated using several techniques. In particular, conjectures based on the U--duality symmetry of the compactified theory have led to proposals for these perturbative contributions to the BPS interactions.    

Our primary aim is to analyze the moduli dependence of these perturbative string amplitudes for the BPS interactions. Rather than directly integrate over the moduli space of the various Riemann surfaces to get the answer which is often quite difficult, we obtain second order differential equations satisfied by these string amplitudes. This $O(d,d,\mathbb{Z})$ invariant differential operator involves only the target space moduli $G_{ij}$ and $B_{ij}$. The various amplitudes satisfy Poisson equations, where the eigenvalues are completely determined by the structure of the integrands in the various multi--loop string amplitudes. The source terms in these equations arise from contributions only from the boundary of the moduli space of Riemann surfaces. This leads to considerable simplification since in order to obtain them, we only need to know the asmyptotic behavior of the integrand at the degenerating and non--degenerating nodes. There are universal source terms which exist in all dimensions, as well as those that exist only in specific dimensions. We determine them directly from the string amplitudes, as well as from U--duality constraints and the field theory limit of these amplitudes where they are encoded in the structure of certain logarithmic ultraviolet divergences in maximal supergravity. This gives the complete differential equations satisfied by these amplitudes in all dimensions. To solve these equations, we need boundary data which is supplied by known results in the decompactification limit. We explicitly obtain these amplitudes in nine and eight dimensions, many of which exist in literature. Particularly significant is the role played in the genus two $D^6\mathcal{R}^4$ amplitude by the Kawazumi--Zhang (KZ) invariant which appears in the integrand. It satisfies a Laplace equation in the interior of moduli space which proves crucial to our analysis. 
Our results prove various conjectures for the differential equations satisfied by these amplitudes, some evidence for which have been provided before.     

We begin by considering the simplest case, namely the type II theory on $S^1$, and obtain and solve the differential equations for the various amplitudes. The primary logic of the analysis generalizes to lower dimensions. We then perform a similar analysis in eight dimensions to obtain explicit solutions for these BPS amplitudes. Finally, we obtain the Poisson equations satisfied by these amplitudes in lower dimensions as well. 

Our primary reason behind pursuing this analysis is that while the amplitudes are difficult to calculate directly by performing the integral over moduli space, it is considerably easier to obtain differential equations satisfied by them. Along with suitable boundary conditions, one can obtain T--duality invariant expressions for them. Our analysis should generalize to string amplitudes in theories with less supersymmetry, where explicit expressions for them are known as integrals over (super)moduli space.

\section{The string amplitudes in nine dimensions}

We first consider the BPS amplitudes in nine dimensions. In this analysis as well as in all the others, we shall make use of the perturbative equality of the four graviton amplitude in the type IIA and IIB theories upto genus three~\cite{Berkovits:2006vc}. This will simplify the solution of the various differential equations to start with. In nine dimensions on compactifying the theory on a circle of radius $r$ in the string frame, this amounts to invariance of the amplitude under interchange of $r$ and $r^{-1}$, while keeping $e^{-2\phi} r$ fixed. This simply follows from the relations~\cite{Hull:1994ys,Witten:1995ex,Aspinwall:1995fw,Schwarz:1995jq}
\be r_A = r_B^{-1}, \quad e^{-\phi_A} = r_B e^{-\phi_B},\ee  
where $\phi_A (\phi_B)$ is the type IIA (IIB) dilaton, and $r_A$ ($r_B$) is the radius of the circle in the type IIA (IIB) theory in the string frame.

At genus zero, the $\mathcal{R}^4$, $D^4\mathcal{R}^4$ and $D^6\mathcal{R}^4$ interactions in nine dimensions have coefficients given by
\be 2\zeta (3) re^{-2\phi}, \quad \zeta(5) re^{-2\phi}, \quad \frac{2}{3} \zeta(3)^2 re^{-2\phi}\ee
respectively.

\subsection{The genus one amplitudes}

We now consider these amplitudes at genus one~\cite{Green:1999pv,Green:2008uj,D'Hoker:2015foa}.
The coefficient of the $\mathcal{R}^4$ interaction is given by
\be \label{1loop1}I^{(1)}_{\mathcal{R}^4} = \pi r \int_{\mathcal{F}_1} \frac{d^2\Omega}{\Omega_2^2} Z_{lat}^{(1)} \equiv\pi r \hat{I}^{(1)}_{\mathcal{R}^4},\ee
while the coefficient of the $D^4\mathcal{R}^4$ interaction is given by
\be \label{1loop2}I^{(1)}_{D^4 \mathcal{R}^4} =\frac{r}{\pi}\int_{\mathcal{F}_1} \frac{d^2\Omega}{\Omega_2^2} Z_{lat}^{(1)} E_2 (\Omega, \bar\Omega) \equiv \frac{r}{\pi} \hat{I}^{(1)}_{D^4\mathcal{R}^4}.\ee
We have defined the $SL(2,\mathbb{Z})$ invariant non--holomorphic Eisenstein series $E_s (\Omega,\bar\Omega)$ as
\bea \label{Eisenstein} E_s(\Omega,\bar\Omega) &&= \sum_{l_i \in \mathbb{Z}, (l_1,l_2) \neq (0,0)} \frac{\Omega_2^s}{\vert l_1 + l_2 \Omega\vert^{2s}} \non \\ && = 2\zeta(2s) \Omega_2^s + 2\sqrt{\pi} \Omega_2^{1-s} \frac{\Gamma(s-1/2)}{\Gamma(s)}\zeta(2s-1)\non \\ && +\frac{4\pi^s \sqrt{\Omega_2}}{\Gamma(s)} \sum_{k\in \mathbb{Z}, k\neq 0} \vert k \vert^{s-1/2} \mu(\vert k\vert, s) K_{s-1/2} (2\pi \Omega_2 \vert k \vert) e^{2\pi i k\Omega_1},\eea
where
\be \mu (k,s) = \sum_{m>0,m|k} \frac{1}{m^{2s-1}}.\ee

Finally, the coefficient of the $D^6\mathcal{R}^4$ interaction is given by
\be \label{1loop3} I^{(1)}_{D^6 \mathcal{R}^4} = \frac{r}{3\pi^2}\int_{\mathcal{F}_1}  \frac{d^2\Omega}{\Omega_2^2} Z_{lat}^{(1)} \Big(5 E_3 (\Omega,\bar\Omega) + \pi^3 \zeta(3)\Big) \equiv \frac{r}{3\pi^2} \hat{I}^{(1)}_{D^6\mathcal{R}^4}+ \frac{\zeta(3)}{3} I^{(1)}_{\mathcal{R}^4}.\ee
In the various expressions we have integrated over $\mathcal{F}_1$, the fundamental domain of $SL(2,\mathbb{Z})$. We follow the convention 
\be d^2\Omega = 2d\Omega_1 d\Omega_2 = d\Omega d\bar\Omega \equiv id\Omega \wedge d\bar\Omega\ee
in defining our measure.
Also the lattice factor $Z_{lat}^{(1)}$ is defined by
\be \label{Lat1}Z_{lat}^{(1)}(\Omega) =  \sum_{m,n \in \mathbb{Z}} e^{-\pi r^2 \vert m+n\Omega\vert^2/\Omega_2}.\ee
Defining the $SL(2,\mathbb{Z})$ invariant Laplacian 
\be \Delta_\Omega = 4\Omega_2^2 \frac{\p^2}{\p\Omega \p\bar\Omega},\ee
we obtain the useful relation
\be \label{lap1}\Delta_\Omega Z_{lat}^{(1)} =  \Big( \lambda^2 \frac{\p^2}{\p\lambda^2} +2 \lambda \frac{\p}{\p\lambda}\Big)Z_{lat}^{(1)}\ee
where $\lambda = r^2$, which relates the Laplacians of the target space and worldsheet moduli. We now simply use \C{lap1} in obtaining the various genus one results. While these results are known, our strategy of calculation easily generalizes to all cases, and this is a warm up exercise. 

For the $\mathcal{R}^4$ interaction, from \C{1loop1} we get that
\be \Big( \lambda^2 \frac{\p^2}{\p\lambda^2} +2 \lambda \frac{\p}{\p\lambda}\Big)\hat{I}^{(1)}_{\mathcal{R}^4} =0,\ee
leading to
\be \label{R4}I^{(1)}_{\mathcal{R}^4} = 4\zeta(2)\Big(r +\frac{1}{r} \Big), \ee
where we have fixed the overall factor using the ten dimensional result. Note that the contribution from the boundary of the moduli space as $\Omega_2 \rightarrow \infty$ involving $\p Z_{lat}^{(1)}/\p\Omega_2$ in the integrand vanishes. 

For the $D^4\mathcal{R}^4$ interaction, from \C{1loop2} we get that
\be \Big( \lambda^2 \frac{\p^2}{\p\lambda^2} +2 \lambda \frac{\p}{\p\lambda}-2\Big)\hat{I}^{(1)}_{D^4\mathcal{R}^4} =0,\ee
leading to
\be \label{DIV1}I^{(1)}_{D^4 \mathcal{R}^4} = \frac{4}{\pi^2}\zeta(3)\zeta(4)\Big(r^3 +\frac{1}{r^3} \Big), \ee
where we have fixed the overall normalization using the answer from two loop supergravity~\cite{Green:1999pu}. This result, which diverges in ten dimensions as $r\rightarrow \infty$, gives rise to non--local terms in the effective action, when added with an infinite number of other such diverging contributions that arise in the decompactification limit. In integrating by parts and obtaining the final answer, we have also used 
\be \label{E}\Delta_\Omega E_s =s(s-1) E_s\ee
and ignored a contribution from the boundary of moduli space $\Omega_2 \rightarrow \infty$ involving 
\be \label{ig1}E_2 \frac{\p Z_{lat}^{(1)}}{\p\Omega_2} - Z_{lat}^{(1)} \frac{\p E_2}{\p\Omega_2} \ee
in the integrand, since it vanishes.

Finally we consider the $D^6\mathcal{R}^4$ interaction in \C{1loop3}. While the second term involving $\zeta(3)$ in the integrand is easily solved using \C{1loop1} and \C{R4}, the first term yields
\be \Big( \lambda^2 \frac{\p^2}{\p\lambda^2} +2 \lambda \frac{\p}{\p\lambda}-6\Big)\hat{I}^{(1)}_{D^6\mathcal{R}^4}=0,\ee
leading to
\be \label{finres}I^{(1)}_{D^6 \mathcal{R}^4} = \frac{10}{\pi^4} \zeta(5)\zeta(6) \Big(r^5 +\frac{1}{r^5}\Big) + \frac{4}{3} \zeta(2)\zeta(3)\Big(r+\frac{1}{r}\Big).\ee
We have again used \C{E} and ignored a vanishing contribution from the boundary of moduli space $\Omega_2 \rightarrow \infty$ involving 
\be \label{ig2}E_3 \frac{\p Z_{lat}^{(1)}}{\p\Omega_2} - Z_{lat}^{(1)} \frac{\p E_3}{\p\Omega_2} \ee
in the integrand. The overall coefficient of the first term in \C{finres} is fixed using the results from two and three loop supergravity, which diverges as $r \rightarrow \infty$~\cite{Green:2005ba,Basu:2014hsa}.

Note that in every case, the $r$ dependence is very easily fixed using \C{lap1} and the perturbative equality of the IIA and IIB theories at this genus. 

\subsection{The genus two amplitudes}

Next we consider the genus two amplitudes of the BPS interactions in nine dimensions~\cite{D'Hoker:2005jc,D'Hoker:2005ht,Berkovits:2005ng}. While the $\mathcal{R}^4$ term vanishes, the coefficient of the $D^4\mathcal{R}^4$ interaction is given by
\be \label{2loop1}I^{(2)}_{D^4\mathcal{R}^4} = \frac{\pi}{2} r e^{2\phi}\int_{\mathcal{F}_2} d\mu_2 Z_{lat}^{(2)} \equiv \frac{\pi}{2} r e^{2\phi} \hat{I}^{(2)}_{D^4\mathcal{R}^4},\ee
while the coefficient of the $D^6\mathcal{R}^4$ interaction is given by
\be \label{2loop2}I^{(2)}_{D^6\mathcal{R}^4} = \pi r e^{2\phi} \int_{\mathcal{F}_2} d\mu_2 Z_{lat}^{(2)} \varphi(\Omega,\bar\Omega) \equiv \pi r e^{2\phi} \hat{I}^{(2)}_{D^6\mathcal{R}^4},\ee
where we have integrated over $\mathcal{F}_2$, the fundamental domain of $Sp(4,\mathbb{Z})$.
Here the $Sp(4,\mathbb{Z})$ invariant measure $d\mu_2$ is given by
\be d\mu_2 = \frac{1}{({\rm det}Y)^3}\prod_{I\leq J} i d\Omega_{IJ} \wedge d\bar\Omega_{IJ},\ee
where $\Omega_{IJ} (I,J=1,2)$ is the period matrix, and $\Omega = X +iY$, where $X$ and $Y$ are matrices with real entries. The fundamental domain $\mathcal{F}_2$ is defined by\footnote{See~\cite{Siegel,Klingen} for various details.}
\bea \vert {\rm Re} \Omega_{IJ} \vert \leq \frac{1}{2}, \quad 0 \leq 2\vert {\rm Im} \Omega_{12} \vert \leq {\rm Im}\Omega_{11} \leq {\rm Im} \Omega_{22}, \quad \vert {\rm det}(C\Omega +D) \geq 1, \eea
where
\be \left( \begin{array}{cc} A & B \\ C & D \end{array} \right)\ee
is an $Sp(4,\mathbb{Z})$ matrix.

The lattice factor is given by
\be \label{Z2}Z_{lat}^{(2)} = \sum_{m_I,n_I} e^{-\pi r^2 (m_I + \Omega_{IK} n_K) Y^{-1}_{IJ}(m_J + \bar\Omega_{JL} n_L)},\ee
where $(Y^{-1})_{IJ} \equiv Y^{-1}_{IJ}$. Finally, $\varphi(\Omega,\bar\Omega)$ is the $Sp(4,\mathbb{Z})$ Kawazumi--Zhang (KZ) invariant~\cite{Zhang,Kawazumi}, which arises in the expression for the genus two amplitude~\cite{D'Hoker:2013eea,D'Hoker:2014gfa}. 

We make repeated use of the useful relation
\be \label{lap2}\Delta_{Sp(4,\mathbb{Z})} Z_{lat}^{(2)} = \Big(\lambda^2 \frac{\p^2}{\p\lambda^2} + 3\lambda \frac{\p}{\p\lambda}\Big) Z_{lat}^{(2)}\ee
in our analysis, where $\Delta_{Sp(4,\mathbb{Z})}$ is the $Sp(4,\mathbb{Z})$ invariant Laplacian on moduli space. 

We first consider the $D^4\mathcal{R}^4$ interaction in \C{2loop1}. Using \C{lap2}, we get that
\be \Big(\lambda^2 \frac{\p^2}{\p\lambda^2} + 3\lambda \frac{\p}{\p\lambda}\Big) \hat{I}^{(2)}_{D^4\mathcal{R}^4}=0,\ee
leading to
\be \label{res}I_{D^4\mathcal{R}^4}^{(2)} = \frac{4}{3}\zeta(4)(e^{-2\phi}r)^{-1} \Big( r^2 +\frac{1}{r^2}\Big),\ee
where we have fixed the overall normalization by using the ten dimensional answer. In the analysis, we have ignored a term of the form $\int_{\mathcal{F}_2} d\mu_2 \Delta_{Sp(4,\mathbb{Z})} Z_{lat}^{(2)}$, which receives contributions only from the boundary of moduli space. To analyze this contribution, we parametrize the period matrix $\Omega$ as
\be \label{parap}\Omega= \left( \begin{array}{cc} \tau_1 & \tau \\ \tau & \tau_2 \end{array} \right).\ee
In this parametrization, the boundary of moduli space involves:\footnote{This analysis generalizes~\cite{D'Hoker:2014gfa} to our case.}

(i) the separating node where the genus two Riemann surface splits into two tori with two additional punctures, 

(ii) the non--separating node where the Riemann surface degenerates to a torus with two additional punctures, 
and

(iii) the intersection of the separating and non--separating nodes. 

The separating node is obtained from \C{parap} by taking $\tau \rightarrow 0$, while keeping $\tau_1, \tau_2$ fixed. The non--separating node is obtained by taking $\tau_2 \rightarrow i\infty$, while keeping $\tau_1, \tau$ fixed. While the boundary term from the non--separating node trivially vanishes from the nature of the lattice sum, the contribution from the separating node also vanishes as one is left with a contour integral $\oint d\tau$ only around the origin, as there is no pole contribution. 

We now consider the $D^6\mathcal{R}^4$ contribution in \C{2loop2}. Again using \C{lap2} we get that
\be \Big( \lambda^2 \frac{\p^2}{\p\lambda^2} + 3\lambda \frac{\p}{\p\lambda}\Big)\hat{I}^{(2)}_{D^6\mathcal{R}^4} = \int_{\mathcal{F}_2} d\mu_2 \varphi (\Omega,\bar\Omega) \Delta_{Sp(4,\mathbb{Z})} Z^{(2)}_{lat}.\ee
On using the relation
\be ({\rm det} Y)^{-3}\Delta_{Sp(4,\mathbb{Z})} = 2\bar\p_{IJ} \Big(({\rm det}Y)^{-3} Y_{IK} Y_{JL} \p_{KL}\Big) +c.c.,\ee
where
\be \p_{IJ} = \frac{1}{2} (1+\delta_{IJ})\frac{\p}{\p\Omega_{IJ}}\ee
and integrating by parts we get that
\be \label{difeqn1}\Big( \lambda^2 \frac{\p^2}{\p\lambda^2} + 3\lambda \frac{\p}{\p\lambda} -5\Big)\hat{I}^{(2)}_{D^6\mathcal{R}^4} =\Psi_1 +\Psi_2,\ee
where we have used the fact that the KZ invariant satisfies the Laplace equation
\be \label{KZ}\Delta_{Sp(4,\mathbb{Z})} \varphi = 5\varphi\ee
in the interior of moduli space.

In \C{difeqn1}, $\Psi_1$ and $\Psi_2$ are contributions localized on the boundary of moduli space and are given by
\bea \label{term}\Psi_1 &=&2\int_{\mathcal{F}_2} d\mu_2 ({\rm det}Y)^3 \Big[ \bar\p_{IJ} \Big(\varphi ({\rm det}Y)^{-3} Y_{IK} Y_{JL} \p_{KL} Z^{(2)}_{lat}\Big)+c.c.\Big],\non \\ \Psi_2 &=& -2\int_{\mathcal{F}_2} d\mu_2 ({\rm det}Y)^3 \Big[ \bar\p_{IJ} \Big(Z^{(2)}_{lat} ({\rm det}Y)^{-3} Y_{IK} Y_{JL} \p_{KL} \varphi\Big)+c.c.\Big].\eea
Hence to evaluate them we only need the behavior of $\varphi$ and $Z^{(2)}_{lat}$ at the boundary of moduli space, leading to considerable simplification. 

Now ${\rm det} Y$ behaves as
\be {\rm det} Y = {\rm Im} \tau_1 {\rm Im} \tau_2 +O(\tau^2)\ee
at the separating node, and as
\be {\rm det} Y = {\rm Im}\tau_1 {\rm Im}\tau_2 + O(\tau_2^0)\ee
at the non--separating node. 

The KZ invariant behaves as~\cite{Wentworth,Jong,D'Hoker:2013eea,Pioline:2015qha}
\be \label{N}\varphi (\Omega, \bar\Omega) = -{\rm ln} \vert 2\pi \tau \eta(\tau_1)^2 \eta(\tau_2)^2\vert + O(\vert \tau\vert^2 {\rm ln} \vert \tau \vert)\ee
at the separating node. At the non--separating node it behaves as
\be \varphi (\Omega,\bar\Omega) = \frac{\pi t}{6} +\hat\varphi_0 (\tau_1,\tau) +\frac{\varphi_1 (\tau_1,\tau)}{t} +O(e^{-t}),\ee
where
\be t= {\rm Im}\tau_2 - \frac{({\rm Im}\tau)^2}{{\rm Im}\tau_1}.\ee
In the expression \C{N}, we have that
\bea \hat\varphi_0 (\tau_1,\tau) = -{\rm ln}\Big[ e^{-\pi({\rm Im}\tau)^2/{\rm Im}\tau_1}\Big\vert \frac{\theta_1 (\tau,\tau_1)}{\eta(\tau_1)}\Big\vert\Big],\non \\ \varphi_1 (\tau_1,\tau)= \frac{5D_{2,2}(\tau_1,\tau)}{16\pi^2}  +\frac{5}{4\pi^3}E_2 (\tau_1,\bar\tau_1).\eea
In the expression for $\varphi_1$, we have 
\be D_{2,2}(\tau_1,\tau) = -\frac{4}{\pi}\sum_{(m,n)\neq (0,0)} \frac{({\rm Im} \tau_1)^2}{\vert m\tau_1 +n\vert^4} e^{\pi[\bar\tau(m\tau_1 +n)- \tau(m\bar\tau_1 +n)]/{\rm Im} \tau_1}\ee
which is a particular case of the Kronecker--Eisenstein series $D_{a,b}(\tau_1,\tau)$~\cite{Zagier}.
Thus on expanding for large ${\rm Im} \tau_2$, we have the expansion
\be \label{nonsep}\varphi(\Omega,\bar\Omega) = \frac{\pi}{6} {\rm Im}\tau_2 + \varphi_0 (\tau_1,\tau)+ \frac{\varphi_1 (\tau_1,\tau)}{{\rm Im}\tau_2}+O(({\rm Im}\tau_2)^{-2})\ee
where
\be \varphi_0 (\tau_1,\tau)=  \frac{5\pi({\rm Im}\tau)^2}{6{\rm Im} \tau_1} -{\rm ln} \Big\vert \frac{\theta_1 (\tau,\tau_1)}{\eta(\tau_1)}\Big\vert.\ee
Finally, the lattice factor $Z^{(2)}_{lat}$ behaves as
\be Z^{(2)}_{lat}= Z^{(1)}_{lat} (\tau_1) Z^{(1)}_{lat} (\tau_2)\ee
at the separating node where $Z^{(1)}_{lat}$ is given by \C{Lat1}, while its behavior 
at the non--separating node is simply obtained by setting ${\rm Im} \tau_2 \rightarrow \infty$.

Now in \C{term} we need to evaluate expressions of the form
\bea &&\bar\p_{IJ}\Big(A ({\rm det}Y)^{-3} Y_{IK} Y_{JL} \p_{KL}  B\Big) + c.c. \non \\ &&= \bar\p_{\tau_1} (A \mathcal{D}_{\tau_1}B) + \bar\p_{\tau_2} (A \mathcal{D}_{\tau_2}B) +\bar\p_{\tau} (A \mathcal{D}_{\tau}B)+c.c.,\eea
where
\bea \mathcal{D}_{\tau_1} &=& ({\rm det}Y)^{-3} Y_{1K} Y_{1L} \p_{KL}, \non \\ \mathcal{D}_{\tau_2} &=& ({\rm det}Y)^{-3} Y_{2K} Y_{2L} \p_{KL}, \non \\ \mathcal{D}_{\tau} &=& ({\rm det}Y)^{-3} Y_{1K} Y_{2L} \p_{KL}.\eea

For both the degenerating nodes, we have that
\bea \label{defD}\mathcal{D}_{\tau_1} &=& ({\rm Im}\tau_1)^{-3}({\rm Im}\tau_2)^{-3}\Big[ ({\rm Im}\tau_1)^2 \p_{\tau_1} + ({\rm Im}\tau)^2 \p_{\tau_2} + {\rm Im}\tau_1{\rm Im}\tau \p_\tau\Big], \non \\ \mathcal{D}_{\tau_2} &=& ({\rm Im}\tau_1)^{-3}({\rm Im}\tau_2)^{-3}\Big[ ({\rm Im}\tau_2)^2 \p_{\tau_2} + ({\rm Im}\tau)^2 \p_{\tau_1} + {\rm Im}\tau_2{\rm Im}\tau \p_\tau\Big], \non \\ \mathcal{D}_{\tau} &=& ({\rm Im}\tau_1)^{-3}({\rm Im}\tau_2)^{-3}\Big[ {\rm Im}\tau_1 {\rm Im}\tau \p_{\tau_1} + {\rm Im}\tau_2 {\rm Im}\tau \p_{\tau_2}\non \\ &&+\frac{1}{2} \Big({\rm Im}\tau_1{\rm Im}\tau_2 +({\rm Im}\tau)^2\Big)\p_\tau\Big],\eea
where we have kept the leading term in ${\rm det}Y$, and kept all the rest. This is all that we need for our analysis. 

Thus we see that at the non--separating node $\tau_2 \rightarrow i \infty$, we have that
\be \mathcal{D}_{\tau_2} \varphi \sim ({\rm Im}\tau_2)^{-1}\ee
and hence such contributions vanish in $\Psi_2$. Also terms involving $\mathcal{D}_{\tau_2} Z^{(2)}_{lat}$ in $\Psi_1$ vanish in this limit and do not yield a finite contribution. Hence there are no contributions from the non--separating node.  

The contribution from the separating node involves a contour integral in $\tau$ around the origin. Hence the contribution from $\varphi \mathcal{D}_\tau Z^{(2)}_{lat}$ in $\Psi_1$ vanishes as there are no poles. On the other hand, from the leading contribution 
\be \mathcal{D}_\tau \varphi  = -\frac{1}{4\tau({\rm Im} \tau_1)^2 ({\rm Im}\tau_2)^2}\ee  
we get a pole term, leading to
\be Z^{(2)}_{lat} \mathcal{D}_\tau \varphi = -\frac{1}{4\tau}\cdot \frac{Z^{(1)}_{lat} (\tau_1)}{({\rm Im}\tau_1)^2}\cdot \frac{Z^{(1)}_{lat} (\tau_2)}{({\rm Im}\tau_2)^2}\ee 
at the separating node. Hence this leads to a non--vanishing contribution from $\Psi_2$\footnote{The contribution from the intersection of the two nodes trivially vanishes as well.}. We get that
\bea \Psi_2 &=& -2 \int_{\mathcal{F}_2} d^2 \tau_1 d^2\tau_2 d(id\tau Z^{(2)}_{lat} \mathcal{D}_\tau \varphi - id\bar\tau Z^{(2)}_{lat} \mathcal{D}_{\bar\tau} \varphi)\non \\ &=& -\frac{1}{2} \int_{\mathcal{F}_1} d^2\tau_1 \int_{\mathcal{F}_1} d^2\tau_2(i \oint d\tau Z^{(2)}_{lat} \mathcal{D}_\tau \varphi - i\oint d\bar\tau Z^{(2)}_{lat} \mathcal{D}_{\bar\tau} \varphi)\non \\ &=& -\frac{\pi}{2} \int_{\mathcal{F}_1} \frac{d^2 \tau_1}{({\rm Im}\tau_1)^2} Z^{(1)}_{lat} (\tau_1)\int_{\mathcal{F}_1} \frac{d^2\tau_2}{({\rm Im}\tau_2)^2} Z^{(1)}_{lat} (\tau_2),\eea
where we have divided by a factor of four in going from the first to the second line. This includes a factor of two because $\tau_1 \leq \tau_2$ in the definition of $\mathcal{F}_2$ whereas we have not kept that ordering in mind, and another factor of two resulting from a discrete $\mathbb{Z}_2$ symmetry from the two tori glued together~\cite{Klingen,Moore:1986rh,D'Hoker:2005ht} in this degeneration limit.
Thus on using \C{1loop1}, we have that
\be \Psi_2 = -\frac{8\zeta(2)^2}{\pi} \Big(1+ \frac{1}{\lambda}\Big)^2,\ee
leading to the differential equation
\be \Big( \lambda^2 \frac{\p^2}{\p\lambda^2} + 3\lambda \frac{\p}{\p\lambda} -5\Big)\hat{I}^{(2)}_{D^6\mathcal{R}^4} = -\frac{8\zeta(2)^2}{\pi} \Big(1+ \frac{1}{\lambda}\Big)^2\ee
for the $D^6\mathcal{R}^4$ amplitude. The solution is
\be \hat{I}^{(2)}_{D^6\mathcal{R}^4} = c_1 r^{2(\sqrt{6}-1)} + c_2 r^{-2(\sqrt{6}+1)} + \frac{8\zeta(2)^2}{5\pi r^2} \Big(r^2 + \frac{1}{r^2} +\frac{5}{3}\Big).\ee
Now $c_1 = c_2=0$ to be consistent with the structure of string perturbation theory, leading to 
\be I^{(2)}_{D^6\mathcal{R}^4} = \frac{8}{5} \zeta(2)^2(e^{-2\phi}r)^{-1}\Big(r^2 + \frac{1}{r^2} +\frac{5}{3}\Big).\ee

\subsection{The genus three amplitude}

Finally we consider the genus three $D^6\mathcal{R}^4$ amplitude in nine dimensions, since the $\mathcal{R}^4$ and $D^4\mathcal{R}^4$ coefficients vanish. The coefficient of this interaction is given by~\cite{Gomez:2013sla}
\be I_{D^6\mathcal{R}^4}^{(3)} = \frac{5}{16} re^{4\phi} \int_{\mathcal{F}_3} d\mu_3 Z_{lat}^{(3)} = \frac{5}{16} r e^{4\phi} \hat{I}^{(3)}_{D^6\mathcal{R}^4},\ee
where $\mathcal{F}_3$ is the fundamental domain of $Sp(6,\mathbb{Z})$, defined after removing some sub--varieties from the Siegel upper half space $\mathcal{S}_3$. Here the $Sp(6,\mathbb{Z})$ invariant measure is given by
\be d\mu_3 = \frac{1}{({\rm det}Y)^4}\prod_{I\leq J} i d \Omega_{IJ} \wedge d\bar\Omega_{IJ},\ee
where $\Omega_{IJ} (I,J=1,2,3)$ is the period matrix, and $\Omega = X +iY$, where $X$ and $Y$ are matrices with real entries. Among other relations, the fundamental domain $\mathcal{F}_3$ is defined by~\cite{Siegel,Klingen}
\be \label{defdom}\vert {\rm Re}\Omega_{IJ} \vert \leq \frac{1}{2},\quad {\rm Im}\Omega_{11} \leq {\rm Im}\Omega_{22} \leq {\rm Im}\Omega_{33}, \quad 2\vert {\rm Im}\Omega_{IJ} \vert \leq {\rm Im}\Omega_{II} ~(1\leq I< J\leq 3),\ee
which will be relevant for our purposes.

The lattice factor is given by
\be \label{Z3}Z_{lat}^{(3)} = \sum_{m_I,n_I} e^{-\pi r^2 (m_I + \Omega_{IK} n_K) Y^{-1}_{IJ}(m_J + \bar\Omega_{JL} n_L)},\ee
where $(Y^{-1})_{IJ} \equiv Y^{-1}_{IJ}$. Making use of the relation
\be \label{lap3}\Delta_{Sp(6,\mathbb{Z})} Z_{lat}^{(3)} = \Big(\lambda^2 \frac{\p^2}{\p\lambda^2} + 4\lambda \frac{\p}{\p\lambda}\Big) Z_{lat}^{(3)},\ee
leads to
\be \Big(\lambda^2 \frac{\p^2}{\p\lambda^2} + 4\lambda \frac{\p}{\p\lambda}\Big) \hat{I}^{(3)}_{D^6\mathcal{R}^4} =0,\ee
which is solved by
\be \label{res2}I_{D^6\mathcal{R}^4}^{(3)} = \frac{4}{27} \zeta(6)(e^{-2\phi} r)^{-2}\Big(r^3 +\frac{1}{r^3}\Big),\ee
where the overall normalization is fixed using the ten dimensional result. As in the genus two case, we have ignored a term of the form $\int_{\mathcal{F}_3} d\mu_3 \Delta_{Sp(6,\mathbb{Z})} Z_{lat}^{(3)}$, which receives contributions only from the boundary of moduli space. The argument is analogous to the genus two analysis. For the non--separating nodes where appropriate worldsheet moduli tend to $i\infty$, the lattice factor causes the boundary contribution to vanish. For the separating nodes where appropriate worldsheet moduli tend to $0$, again the contributions vanish due to the lack of pole terms in the contour integrals. Our results agree with results based on U--duality~\cite{Basu:2007ru,Basu:2007ck,Green:2010wi,Green:2010kv}.

In the above analysis, we have been quite cavalier about ignoring various total derivatives on moduli space. While we could do so in nine dimensions, this is no longer true in general as we shall see later.  

\section{The string amplitudes in eight dimensions}

We now consider these BPS amplitudes in eight dimensions, where the analysis gets more involved. The T--duality group is $SL(2,\mathbb{Z})_T \times SL(2,\mathbb{Z})_U$, where $T$ and $U$ are the Kahler and complex structure moduli of the $T^2$. Here
\be T = B_{12} +iV_2,\ee
where $B_{12}$ is the NS--NS sector modulus coming from the 2--form in ten dimensions and $V_2$ is the volume of the $T^2$ in the string frame. Perturbative equality of the IIA and IIB amplitudes implies invariance under interchange of $T$ and $U$ while keeping $e^{-2\phi}T_2$ fixed. 

At tree level, the $\mathcal{R}^4$, $D^4 \mathcal{R}^4$ and $D^6\mathcal{R}^4$ interactions have coefficients
\be 2\zeta(3) T_2 e^{-2\phi}, \quad \zeta(5) T_2 e^{-2\phi}, \quad \frac{2}{3} \zeta(3)^2 T_2 e^{-2\phi}\ee 
respectively.

\subsection{The genus one amplitudes}

We write down the expressions for the genus one amplitudes for compactifications on $T^d$ for arbitrary $d$, as we shall find them useful later.

The coefficient of the $\mathcal{R}^4$ interaction is given by
\be \label{1genus1}I^{(1)}_{\mathcal{R}^4} = \pi  \int_{\mathcal{F}_1} \frac{d^2\Omega}{\Omega_2^2} \Gamma_{d,d;1} ,\ee
while the coefficient of the $D^4\mathcal{R}^4$ interaction is given by 
\be \label{1genus2}I^{(1)}_{D^4 \mathcal{R}^4} =\frac{1}{\pi}\int_{\mathcal{F}_1} \frac{d^2\Omega}{\Omega_2^2}  E_2 (\Omega, \bar\Omega)\Gamma_{d,d;1}.\ee
Finally the coefficient of the $D^6\mathcal{R}^4$ interaction is given by
\be \label{1genus3} I^{(1)}_{D^6 \mathcal{R}^4} = \frac{1}{3\pi^2}\int_{\mathcal{F}_1}  \frac{d^2\Omega}{\Omega_2^2} \Big(5 E_3 (\Omega,\bar\Omega) + \pi^3 \zeta(3)\Big) \Gamma_{d,d;1}= \hat{I}^{(1)}_{D^6\mathcal{R}^4}+\frac{\zeta(3)}{3} I^{(1)}_{\mathcal{R}^4}. \ee

In the above expressions, the lattice factor at genus $g$ for compactification on $T^d$ is given by\footnote{Thus, $\Gamma_{1,1;g}(r,\Omega) = r^g Z^{(g)}_{lat}$, for $Z^{(g)}_{lat}$ given by \C{Lat1}, \C{Z2} and \C{Z3}.}
\be \label{genlat}\Gamma_{d,d;g} (G,B;\Omega)= (V_d)^g \sum_{m_I^i,n_I^i\in \mathbb{Z}} e^{-\pi(G+B)_{ij}(m_I^i +\Omega_{IK}n_K^i)Y^{-1}_{IJ}(m_J^j +\bar\Omega_{JL}n_L^j)},\ee
where $I,J=1,\ldots,g$ and $i,j=1,\ldots,d$.

Thus in eight dimensions, the lattice factor at genus one is given by $\Gamma_{2,2;1}$,
where the metric of the target space $T^2$ is given by
\be G_{ij}= \frac{T_2}{U_2}\left( \begin{array}{cc} 1 & U_1 \\ U_1 & \vert U\vert^2 \end{array} \right).\ee
We make repeated use of the relation~\cite{Obers:1999um}
\be 2\Delta_\Omega \Gamma_{2,2;1} = (\Delta_T +\Delta_U)\Gamma_{2,2;1}\ee
relating the various Laplacians.

Thus for the $\mathcal{R}^4$ interaction in \C{1genus1} we see that
\be \label{div}(\Delta_T + \Delta_U) I^{(1)}_{\mathcal{R}^4} = 2\pi\int_{\mathcal{F}_1^L} \frac{d^2\Omega}{\Omega_2^2} \Delta_\Omega \Gamma_{2,2;1} = 4\pi \int_{-1/2}^{1/2} d\Omega_1 \frac{\p\Gamma_{2,2;1}}{\p\Omega_2}\Big\vert_{\Omega_2 =L \rightarrow \infty}, \ee
where we have integrated over the truncated part of $\mathcal{F}_1$ such that $\Omega_2 \leq L$~\cite{Green:1999pv,Green:2008uj}. This is done to isolate the contribution to the analytic part of the amplitude from the part of the amplitude which is non--analytic in the external momenta. The former receives contributions from the part of the moduli space with $\Omega_2 \leq L$, and the later receives contributions only from $\Omega_2 > L$. At the end of the calculation we have to take $L\rightarrow \infty$. The constant term in the integral over $\mathcal{F}_1^L$ is the analytic part of the amplitude, while the $L$ dependent divergent parts cancel from the part of the amplitude for $\Omega_2 > L$. Now the right hand side of \C{div} involves an integral over the boundary of moduli space, which inspite of being a total derivative, does not vanish due to infrared effects. For the genus one $\mathcal{R}^4$ amplitudes in various dimensions, this is only true in eight dimensions, as will be clear from our analysis below.  

To see this, we use an alternative representation for the lattice sum \C{genlat} given by
\be \label{altgenlat}\Gamma_{d,d;g} (G,B;\Omega)= ({\rm det}Y)^{d/2} \sum_{m_I^i, n_I^i \in \mathbb{Z}} e^{-\pi Y_{IJ} \mathcal{L}_{IJ} +2\pi i m_I^i X_{IJ} n_J^i},\ee   
where
\be \mathcal{L}_{IJ} = G^{ij} (m_I^i + B_{ik} n_I^k)(m_J^j + B_{jl} n_J^l) + G_{ij} n_I^i n_J^j.\ee
This representation has the advantage of making the boundary behavior of relevant expressions simpler, in particular its divergences. In fact, to obtain the representation \C{genlat} one Poisson resums over all momentum modes to express the lattice sum only in terms of winding modes. The expression \C{altgenlat} on the other hand, involves a sum over both momentum and winding modes.      

To evaluate \C{div} we see that the only non--vanishing contribution comes when all the integers in the lattice sum \C{altgenlat} vanish leading to
\be \label{deqn}(\Delta_T + \Delta_U) I^{(1)}_{\mathcal{R}^4}  = 4\pi.\ee
The entire contribution thus comes from $\p/\p\Omega_2$ acting on $\Omega_2^{d/2}$ in $\Gamma_{d,d;1}$ which yields a constant only for $d=2$. Thus such boundary contributions vanish in all other dimensions.    

We now solve \C{deqn} to obtain\footnote{In the cases where we solve the Poisson equations explicitly in eight dimensions, we assume that cusp forms do not contribute to the answer. These are exponentially supressed in the decompactification limit and hence are not constrained by decompactification. In several cases, they are indeed absent which follows from a direct evaluation of the amplitude. }
\be \label{readoff}I^{(1)}_{\mathcal{R}^4} = 2\Big(E_1 (T,\bar{T}) +E_1 (U,\bar{U})\Big),\ee 
where\footnote{This is defined after subtracting a pole.}
\be E_1 (T,\bar{T}) = -\pi{\rm ln}(T_2 \vert \eta(T)\vert^4).\ee

In fact, one can evaluate the integral directly in \C{1genus1}, leading to\footnote{Several regularizations have been used to evaluate it. This includes the method of orbits~\cite{Dixon:1990pc}, or evaluating it with $-E_s (\Omega,\bar\Omega)$ inserted in the integrand, and sending $s\rightarrow 0$ at the end~\cite{Green:2010wi,Pioline:2014bra}. All of them involve the unfolding technique, and yield the same answer upto a moduli independent constant which can be absorbed in the scale of the logarithm in the final answer.} \C{readoff}. Our derivation is simpler because it reduces the entire computation to a boundary term, which is much easier to evaluate than the whole integral. In particular, the unfolding technique can be avoided.

While this is known, we would like to derive \C{deqn} from considerations of U--duality, as this will be very useful for our purposes when generalizing to certain higher genus amplitudes in lower dimensions. The eight dimensional effective action has a non--analytic term of the schematic form
\be \int d^8 x \sqrt{-g^{(8)}} {\rm ln}(-\mu \alpha' s) \mathcal{R}^4\ee
in the string frame at genus one, where $\mu$ is a constant, and $s$ is a generic Mandelstam variable. This is obtained from the logarithmic ultraviolet divergence of one loop supergravity~\cite{Green:1982sw,Russo:1997mk,Green:2010sp}, which produces an infrared divergence in string theory. On converting to the Einstein frame, it yields a dilaton dependent local contribution given by
\be \label{scale1}\frac{4\pi}{3} {\rm ln} g_2\ee 
to the $\mathcal{R}^4$ interaction, where $g_d^{-2} = e^{-2\phi} V_d \equiv e^{-2\phi_d}$. This is a part of the perturbative expansion of the $SL(3,\mathbb{Z}) \times SL(2,\mathbb{Z})$ invariant $\mathcal{R}^4$ coefficient function $\mathcal{E}_{\mathcal{R}^4}$ which satisfies~\cite{Kiritsis:1997em,Green:2010wi,Green:2010kv}
\be \label{VAn}(\Delta_{\mathcal{U}} +\ldots )\mathcal{E}_{\mathcal{R}^4} = 6\pi,\ee 
where $\Delta_{\mathcal{U}}$ is the U--duality invariant Laplacian. Now
\be \label{arbd}\Delta_{\mathcal{U}} = \frac{8-d}{8} \p^2_{\phi_d}+\frac{d^2-d+4}{4} \p_{\phi_d}+\Delta_{O(d,d,\mathbb{Z})} +\ldots,\ee
where $\Delta_{O(d,d,\mathbb{Z})}$ is the $O(d,d,\mathbb{Z})$ invariant Laplacian. This leads to the $4\pi$ in \C{deqn} at genus one on converting to the string frame\footnote{We have that $\Delta_{O(2,2,\mathbb{Z})} = \Delta_T +\Delta_U$.}. The eigenvalue in equation \C{VAn} vanishes in eight dimensions, hence the answer is independent of the scale of the logarithm in \C{scale1}. Thus we see that this coefficient, which appears only in eight dimensions, is entirely fixed by the structure of the logarithmic divergence of the one loop non--local supergravity amplitude. We shall see the generalization of this analysis to higher genus amplitudes later.

For the $D^4\mathcal{R}^4$ interaction from \C{1genus2}, we have 
\be (\Delta_T + \Delta_U -4)I^{(1)}_{D^4\mathcal{R}^4} =0,\ee
which is solved by
\be \label{1g1}I^{(1)}_{D^4\mathcal{R}^4} =\frac{2}{\pi^3}E_2 (T,\bar{T}) E_2 (U,\bar{U})\ee
where we have fixed the overall constant by matching with the decompactification limit\footnote{We often need to find solutions of the homogeneous differential equation
\be (\Delta_T +\Delta_U -\lambda)I=0.\ee
Consistent with the structure of the perturbative equality of the type IIA and IIB amplitudes, the solutions are
\be I\sim E_s(T,\bar{T})E_s (U,\bar{U}),\ee
where $\lambda = 2s(s-1)$ and
\be I\sim E_s(T,\bar{T})+E_s(U,\bar{U}),\ee
where $\lambda =s(s-1)$. We consider only the ones that might be allowed in string theory.}. 
Note that we have used \C{E} and dropped a boundary contribution of the form \C{ig1} involving $\Gamma_{2,2;1}$.

Now let us consider the $D^6\mathcal{R}^4$ interaction in \C{1genus3}. While the contribution from the second term follows from \C{1genus1} and \C{readoff}, the first term is given by\footnote{A solution of the form $E_4 (T,\bar{T}) +E_4 (U,\bar{U})$ is inconsistent as it behaves as $T_2^4$ for large $T_2$.}
\be (\Delta_T + \Delta_U -12)\hat{I}^{(1)}_{D^6\mathcal{R}^4} =0,\ee
leading to 
\be \label{1g2}I^{(1)}_{D^6 \mathcal{R}^4} =\frac{20}{3\pi^5} E_3 (T,\bar{T})E_3 (U,\bar{U}) + \frac{2}{3}\zeta(3)\Big(E_1 (T,\bar{T}) + E_1 (U,\bar{U})\Big),\ee
where we have fixed the coefficient of the first term by matching with the decompactification limit. 
We have again used \C{E} and dropped a boundary contribution of the form \C{ig2} involving $\Gamma_{2,2;1}$.

\subsection{The genus two amplitudes}

On compactifying on $T^d$ at genus two, the $D^4\mathcal{R}^4$ interaction is given by
\be \label{1Genus2}I^{(2)}_{D^4\mathcal{R}^4} = \frac{\pi}{2}(e^{-2\phi}V_d)^{-1} \int_{\mathcal{F}_2} d\mu_2 \Gamma_{d,d;2}=\frac{\pi}{2}(e^{-2\phi}V_d)^{-1} \hat{I}^{(2)}_{D^4\mathcal{R}^4} ,\ee
while the $D^6\mathcal{R}^4$ interaction is given by
\be \label{2Genus2}I^{(2)}_{D^6\mathcal{R}^4} = \pi(e^{-2\phi}V_d)^{-1} \int_{\mathcal{F}_2} d\mu_2 \varphi (\Omega,\bar\Omega)\Gamma_{d,d;2}=\pi(e^{-2\phi}V_d)^{-1} \hat{I}^{(2)}_{D^6\mathcal{R}^4} .\ee
We make use of the relation~\cite{Obers:1999um}
\be (\Delta_T +\Delta_U -2) \Gamma_{2,2;2} = 2\Delta_{Sp(4,\mathbb{Z})} \Gamma_{2,2;2}\ee
in our analysis. Thus the $D^4\mathcal{R}^4$ amplitude satisfies
\be (\Delta_T +\Delta_U -2) \hat{I}^{(2)}_{D^4\mathcal{R}^4} = 0,\ee
on dropping boundary contributions which vanish. This leads to
\be I^{(2)}_{D^4\mathcal{R}^4} = \frac{2}{3}(e^{-2\phi}T_2)^{-1}\Big( E_2(T,\bar{T})+E_2 (U,\bar{U})\Big),\ee
where we have fixed the overall constant by considering the decompactification limit. 

For the $D^6\mathcal{R}^4$ interaction, 
this leads to
\be \label{dim8}(\Delta_T +\Delta_U -12) \hat{I}^{(2)}_{D^6\mathcal{R}^4} = 2(\Psi_1 +\Psi_2),\ee
where we have integrated by parts, and used \C{KZ}. In \C{dim8}, $\Psi_1$ and $\Psi_2$ are given by \C{term} with $Z^{(2)}_{lat}$ replaced by $\Gamma_{2,2;2}$. The boundary contributions can be analyzed exactly as in the nine dimensional case, and only $\Psi_2$ receives a non--vanishing contribution from the separating node, leading to
\be 2 \Psi_2 = -\pi \int_{\mathcal{F}_1} \frac{d^2 \tau_1}{({\rm Im}\tau_1)^2}  \Gamma_{2,2;1}(\tau_1)\int_{\mathcal{F}_1} \frac{d^2\tau_2}{({\rm Im}\tau_2)^2}  \Gamma_{2,2;1}(\tau_2).\ee 
Thus on using \C{1genus1} and \C{readoff}, we get that
\be \pi\Big(\Delta_T +\Delta_U -12\Big) \hat{I}^{(2)}_{D^6\mathcal{R}^4} = -4 \Big(E_1 (T,\bar{T}) + E_1 (U,\bar{U})\Big)^2.\ee
Note that \C{KZ} is crucial in obtaining the eigenvalue 12.

\begin{figure}[ht]
\begin{center}
\[
\mbox{\begin{picture}(140,70)(0,0)
\includegraphics[scale=.4]{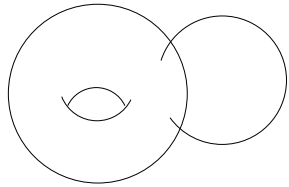}
\end{picture}}
\]
\caption{The non--separating genus two node}
\end{center}
\end{figure}

There is one further contribution that we need to include to complete the analysis\footnote{This is a contribution from the non--separating node. See the discussion following \C{van} for details.}. This comes from the $O({\rm Im}\tau_2)$ term in the expression for the KZ invariant $\varphi$ in \C{nonsep}. This is the contribution from the part of the moduli space where both ${\rm Im}\tau_1 \rightarrow \infty$ and ${\rm Im}\tau_2 \rightarrow \infty$. This is the field theory limit of the amplitude\footnote{This limit corresponds to particle propagating over infinite proper time, hence only massless modes propagate.}, as it corresponds to the limit where the entire Riemann surface reduces to the two loop skeleton diagram of the supergravity amplitude. This contribution arises from a suitable part of the contribution from the non--separating node (depicted by figure 1) which localizes at ${\rm Im}\tau_2\rightarrow \infty$\footnote{Other contributions from the non--separating node are discussed in the next section. They vanish in eight dimensions.}. Thus in the expression for $\mathcal{D}_{\tau_2}$ in \C{defD} we choose the term involving $\p/\p\tau_1$ to obtain       
\be \frac{\pi}{18({\rm Im}\tau_2)^{2-d/2}} \int_{\mathcal{F}_1} d^2 \tau_1 \frac{\p}{\p{\rm Im}\tau_1}\Gamma_{d,d;1} (\tau_1) \Big\vert_{{\rm Im}\tau_2 \rightarrow \infty} \ee
where we have integrated over $\tau$. This receives non--vanishing contribution only from the boundary at ${\rm Im}\tau_1\rightarrow\infty$ which is precisely the field theory limit, to yield
\be  \frac{\pi({\rm Im}\tau_1)^{d/2}}{9 ({\rm Im}\tau_2)^{2-d/2}}\Big\vert_{{\rm Im}\tau_i \rightarrow \infty}.\ee  
For the special case when $d=2$, this reduces to
\be \frac{\pi}{9} \Big( \frac{{\rm Im}\tau_1}{{\rm Im}\tau_2}\Big)\Big\vert_{{\rm Im}\tau_i \rightarrow \infty} = {\rm constant},\ee
since ${\rm Im}\tau_1/{\rm Im}\tau_2 \sim O(1)$. Hence we have that\footnote{Such field theory contributions arising from the boundary of moduli space have been directly evaluated for the various cases in~\cite{Pioline:2015nfa} by explicit calculation.}
\be \label{maineqn}\pi\Big(\Delta_T +\Delta_U -12\Big) \hat{I}^{(2)}_{D^6\mathcal{R}^4} = -4 \Big(E_1 (T,\bar{T}) + E_1 (U,\bar{U})\Big)^2 +\hat{c} ,\ee
where $\hat{c}$ is a constant. Making the ansatz 
\be \label{ansatz}\pi\hat{I}^{(2)}_{D^6\mathcal{R}^4}  = f(T,\bar{T}) +f(U,\bar{U}) + \alpha\Big(E_1 (T,\bar{T}) +E_1 (U,\bar{U})\Big) +\beta E_1 (T,\bar{T}) E_1 (U,\bar{U}) +\gamma,\ee
where $\alpha$, $\beta$ and $\gamma$ are constants, from \C{maineqn} we get that
\be \label{ansatz2}\alpha = \frac{\pi}{18}, \quad \beta = \frac{2}{3}, \quad \hat{c} = -12\gamma +\frac{\pi^2}{9},\ee
while $f(T,\bar{T})$ satisfies
\be (\Delta_T -12) f(T,\bar{T}) = -4 E_1 (T,\bar{T})^2.\ee
Note that we did not add 
\be c_1\Big(E_4 (T,\bar{T}) +E_4 (U,\bar{U})\Big), \ee
a solution to the homogeneous equation \C{maineqn} to the ansatz \C{ansatz}, since $E_4 (T,\bar{T}) \sim T_2^4$ for large $T_2$ which is inconsistent. Also we did not add\be \label{no}c_2 E_3 (T,\bar{T}) E_3 (U,\bar{U}),\ee 
another solution to the homogeneous equation \C{maineqn} to the ansatz \C{ansatz}. This follows from a feature of the decompactification limit of the answer to nine dimensions on analyzing non--local contributions~\cite{Green:2006gt}. To see this, we decompactify to nine dimensions, using
\be T_2 = r_\infty r, \quad U_2 = \frac{r_\infty}{r},\ee
where we take $r_\infty \rightarrow \infty$. Setting $l_s \int d^8 x \sqrt{-g^{(8)}} r_\infty = \int d^9 x \sqrt{-g^{(9)}}$, we see that the genus one $D^4\mathcal{R}^4$ and $D^6\mathcal{R}^4$ amplitudes \C{1g1} and \C{1g2} produce contributions $16\zeta(4)^2 \pi^{-3} r_\infty^3$ and $40 \zeta(6)^2 \pi^{-5} r_\infty^5$ respectively, which diverge as $r_\infty \rightarrow \infty$. These are the leading contributions to the infinite sum of the form\footnote{For decompactification to ten dimensions on using $l_s \int d^9 x \sqrt{-g^{(9)}} r = \int d^{10} x \sqrt{-g^{(10)}}$, the genus one $D^4\mathcal{R}^4$ and $D^6\mathcal{R}^4$ contributions \C{DIV1} and \C{finres} which diverge are $8\zeta(3)\zeta(4)\pi^{-2} r^2$ and $15\zeta(5)\zeta(6)\pi^{-5}r^4$ respectively, which are the leading contributions to the sum
\be r^{-2}\sum_{n=2}^\infty b_n (r^2 s)^n \mathcal{R}^4,\ee
$b_n$ being constants. This produces the threshold correction $s{\rm ln}(-\alpha' s)\mathcal{R}^4$.}
\be r^{-1}_\infty \sum_{n=2}^\infty a_n (r_\infty^2 s)^n \mathcal{R}^4\ee
for constant $a_n$ which add up to produce threshold corrections of the form $(-\alpha' s)^{1/2} \mathcal{R}^4$ in nine dimensions. However, \C{no} produces a divergence $(e^{-2\phi} r)^{-1} 4\zeta(6)^2 r_\infty^4$ in nine dimensions which does not lead to an admissible threshold correction, hence $c_2 =0$.

We shall now fix $\gamma$ and $\hat{c}$ using the constraints imposed by U--duality. The eight dimensional effective action has a non--analytic term of the form
~\cite{Bern:1998ug,Green:2010sp}
\be l_s^6 \int d^8 x \sqrt{-g^{(8)}} (e^{-2\phi} V_2)^{-1}{\rm ln}(-\mu \alpha' s) D^6\mathcal{R}^4\ee
in the string frame at genus two. This is obtained from the logarithmic ultraviolet divergence of two loop supergravity. This arises from the two loop contribution as well as the contribution coming from a one loop counterterm. On converting to the Einstein frame, it yields a dilaton dependent local contribution given by
\be \label{scheme}\frac{4\pi^2}{27} {\rm ln}^2 g_2 +\frac{2\pi}{9}\Big[\frac{\pi}{2} + 2\Big(E_1 (T,\bar{T}) +E_1 (U,\bar{U})\Big)\Big] {\rm ln} g_2 \ee
to the $D^6\mathcal{R}^4$ interaction. This is a part of the perturbative expansion of the $SL(2,\mathbb{Z}) \times SL(3,\mathbb{Z})$ invariant $D^6 \mathcal{R}^4$ coefficient function $\mathcal{E}_{D^6\mathcal{R}^4}$ which satisfies
\be (\Delta_\mathcal{U} +\ldots )\mathcal{E}_{D^4 \mathcal{R}^4} = 0,\ee 
where $\Delta_\mathcal{U}$ is the $SL(2,\mathbb{Z})\times SL(3,\mathbb{Z})$ invariant Laplacian\footnote{The eigenvalue does not vanish in eight dimensions.}.
In obtaining \C{scheme}, a specific scheme of choosing the scale of the logarithm has been used~\cite{Green:2010wi}. In converting from the string frame to the Einstein frame, only the dilaton dependent part is included in the definition of the local action, while the entire $\mu$ dependence is kept in the definition of the non--local term. Thus the term
\be {\rm ln}(-\alpha' e^{\lambda\phi} \mu s) D^{2p} \mathcal{R}^4\ee 
contributes $\lambda\phi D^{2p}\mathcal{R}^4$ to the local action, and ${\rm ln}(-\alpha' \mu s)D^{2p}\mathcal{R}^4$ to the non--local action, for example. In this scheme of regularization, the complete genus two contribution in the Einstein frame is given by~\cite{Green:2010wi}
\bea \label{ansatz3}f(T,\bar{T}) +f(U,\bar{U}) + \frac{\pi}{9}\Big(E_1 (T,\bar{T}) +E_1 (U,\bar{U})\Big) +\frac{2}{3} E_1 (T,\bar{T}) E_1 (U,\bar{U}) \non \\ +\frac{\pi}{18} \Big[\pi +4\Big(E_1 (T,\bar{T}) +E_1 (U,\bar{U})\Big)\Big]{\rm ln} g_2^2 +\frac{2\zeta(2)}{9} ({\rm ln} g_2^2)^2  +\frac{11\zeta(2)}{36}.\eea
We now compare \C{ansatz} and \C{ansatz2} with \C{ansatz3}. We use the freedom of rescaling the logarithm to send 
\be \label{rescale}{\rm ln} g_2^2 \rightarrow {\rm ln} (\mu g_2^2),\ee 
where
\be {\rm ln}\mu = -\frac{1}{4}\ee
thus equating the coefficient of the $E_1 (T,\bar{T}) +E_1 (U,\bar{U})$ term which is independent of the logarithm in \C{ansatz} and \C{ansatz3}\footnote{Equivalently we fix the ambiguity of the scale of the logarithm differently. These correspond to different renormalization schemes.}. Then \C{ansatz3} goes to
\bea \label{final}f(T,\bar{T}) +f(U,\bar{U}) + \frac{\pi}{18}\Big(E_1 (T,\bar{T}) +E_1 (U,\bar{U})\Big) +\frac{2}{3} E_1 (T,\bar{T}) E_1 (U,\bar{U}) \non \\ +\frac{\pi}{9} \Big[\frac{\pi}{3} +2\Big(E_1 (T,\bar{T}) +E_1 (U,\bar{U})\Big)\Big]{\rm ln} g_2^2 +\frac{2\zeta(2)}{9} ({\rm ln} g_2^2)^2  +\frac{17\zeta(2)}{72}.\eea  
Hence in this scheme of renormalization, we compare \C{ansatz} and \C{final} leading to
\be \gamma = \frac{17\zeta(2)}{72}, \quad \hat{c} = -\frac{13\zeta(2)}{6}.\ee
In fact, before the rescaling \C{rescale}, the structure of the differential equation satisfied by the local part of the genus two amplitude in the string frame is different in the two analyses. While we have \C{maineqn}, the one following from \C{ansatz3} has extra source terms 
\be -\frac{\pi}{3}I^{(1)}_{\mathcal{R}^4} - \frac{7\pi^2}{18}.\ee
However we see that a potential source term linear in $I^{(1)}_{\mathcal{R}^4}$ cancels (see the arguments following \C{van}) in the calculation of the string amplitude. Hence our choice of assigning the scale of the logarithm is the natural one from the point of view of the genus two amplitude.

\subsection{The genus three amplitude}

On compactifying on $T^d$ at genus three, the $D^6\mathcal{R}^4$ interaction is given by
\be \label{G3}I^{(3)}_{D^6\mathcal{R}^4} = \frac{5}{16}(e^{-2\phi}V_d)^{-2} \int_{\mathcal{F}_3} d\mu_3 \Gamma_{d,d;3}=\frac{5}{16}(e^{-2\phi}V_d)^{-2} \hat{I}^{(2)}_{D^6\mathcal{R}^4} .\ee
Note that in eight dimensions~\cite{Obers:1999um}
\be (\Delta_T +\Delta_U -6) \Gamma_{(2,2;3)} = 2 \Delta_{Sp(6,\mathbb{Z})} \Gamma_{2,2;3}.\ee
Thus
\be (\Delta_T +\Delta_U -6) \hat{I}^{(2)}_{D^6\mathcal{R}^4} = 0,\ee
on dropping boundary contributions which vanish. This leads to
\be I^{(3)}_{D^4\mathcal{R}^4} = \frac{2}{27}\Big(E_3(T,\bar{T})+E_3 (U,\bar{U})\Big),\ee
where we have fixed the overall constant by considering the decompactification limit.
 These results are consistent with the U--duality analysis of~\cite{Basu:2007ru,Basu:2007ck,Green:2010wi,Green:2010kv}.

\section{The structure of the equations in general dimensions}

We now consider the general structure of the differential equations satisfied by the $\mathcal{R}^4$, $D^4\mathcal{R}^4$ and $D^6\mathcal{R}^4$ amplitudes in compactifications on $T^d$ for $d \leq 7$. We shall write down differential equations satisfied by these amplitudes, but not solve them explicitly like we did in nine and eight dimensions. They can be solved along the lines of the earlier analysis, with the boundary conditions provided by the various decompactification limits\footnote{Apart from possible cusp forms, which need a separate analysis.}. Some of the details are very similar to what we have already done.  

\subsection{The genus one amplitudes}

At genus one, the coefficients of the $\mathcal{R}^4$, $D^4\mathcal{R}^4$ and $D^6\mathcal{R}^4$ interactions are given by \C{1genus1}, \C{1genus2} and \C{1genus3} respectively.

We repeatedly make use of the relation~\cite{Obers:1999um}
\be \Big(\Delta_{O(d,d,\mathbb{Z})} +\frac{d(d-2)}{2}\Big)\Gamma_{d,d;1} = 2\Delta_\Omega \Gamma_{d,d;1}\ee
to obtain the various differential equations. For the $\mathcal{R}^4$ amplitude, we get that
\be \Big(\Delta_{O(d,d,\mathbb{Z})} +\frac{d(d-2)}{2}\Big) I^{(1)}_{\mathcal{R}^4} =4\pi \delta_{d,2}, \ee  
where the origin of the source term has been explained before. For the $D^4\mathcal{R}^4$ amplitude, we get that
\be\label{NV1}\Big(\Delta_{O(d,d,\mathbb{Z})} +\frac{(d+2)(d-4)}{2}\Big) I^{(1)}_{D^4 \mathcal{R}^4} =\frac{4}{\pi} \int_{-1/2}^{1/2} d\Omega_1  \Big( E_2 \frac{\p\Gamma_{d,d;1}}{\p\Omega_2} -\Gamma_{d,d;1} \frac{\p E_2}{\p\Omega_2}\Big) \Big\vert_{\Omega_2 = L\rightarrow \infty}. \ee
Using
\be E_2 (\Omega,\bar\Omega) = 2\zeta(4)\Omega_2^2 +\frac{\pi\zeta(3)}{\Omega_2}+\ldots\ee
on neglecting terms which vanish exponentially as $\Omega_2 \rightarrow \infty$, we see that \C{NV1} can yield a non--vanishing contribution only for $d=4$ on using \C{altgenlat}. Thus, we get that
\be \Big(\Delta_{O(d,d,\mathbb{Z})} +\frac{(d+2)(d-4)}{2}\Big) I^{(1)}_{D^4 \mathcal{R}^4}  = 12\zeta(3) \delta_{d,4}.\ee

Finally for the $D^6\mathcal{R}^4$ amplitude we have that 
\be \label{NV2}\Big(\Delta_{O(d,d,\mathbb{Z})} +\frac{(d+4)(d-6)}{2}\Big) \hat{I}^{(1)}_{D^6 \mathcal{R}^4} =\frac{20}{3\pi^2} \int_{-1/2}^{1/2} d\Omega_1  \Big( E_3 \frac{\p\Gamma_{d,d;1}}{\p\Omega_2} -\Gamma_{d,d;1} \frac{\p E_3}{\p\Omega_2}\Big) \Big\vert_{\Omega_2 = L\rightarrow \infty}.\ee
Using
\be E_3 (\Omega,\bar\Omega) = 2\zeta(6)\Omega_2^3 +\frac{3\pi\zeta(5)}{4\Omega_2^2}+\ldots\ee
on neglecting terms which vanish exponentially as $\Omega_2 \rightarrow \infty$, we see that \C{NV2} can yield a non--vanishing contribution only for $d=6$ leading to
\be \Big(\Delta_{O(d,d,\mathbb{Z})} +\frac{(d+4)(d-6)}{2}\Big) \hat{I}^{(1)}_{D^6 \mathcal{R}^4} = \frac{25\zeta(5)}{\pi} \delta_{d,6}.\ee

\subsection{The genus two amplitudes}

At genus two, the $D^4\mathcal{R}^4$ and $D^6\mathcal{R}^4$ interactions are given by \C{1Genus2} and \C{2Genus2} respectively.

We make use of the relation~\cite{Obers:1999um} 
\be \Big(\Delta_{O(d,d,\mathbb{Z})} +d(d-3)\Big) \Gamma_{d,d;2} = 2\Delta_{Sp(4,\mathbb{Z})} \Gamma_{d,d;2},\ee
so that the $D^4\mathcal{R}^4$ amplitude satisfies the equation
\be \label{Arb}\Big(\Delta_{O(d,d,\mathbb{Z})} +d(d-3)\Big) \hat{I}^{(2)}_{D^4\mathcal{R}^4} = 2\int_{\mathcal{F}_2} d\mu_2 \Delta_{Sp(4,\mathbb{Z})}\Gamma_{d,d;2}, \ee
yielding a boundary term on the right hand side. The contribution from the separating node vanishes as discussed before, due to the lack of a pole term in the $\oint d\tau$ integral. The contribution from the non--separating node to the right hand side of \C{Arb} is given by  
\be 4 \int_{\mathcal{F}_2} d\mu_2 ({\rm det} Y)^3 (\bar\p_{\tau_2} \mathcal{D}_{\tau_2} + \p_{\tau_2}  \bar{\mathcal{D}}_{\tau_2})\Gamma_{d,d;2}\ee
in the limit ${\rm Im}\tau_2 \rightarrow \infty$. Using the first term in the expression for $\mathcal{D}_{\tau_2}$ in \C{defD}, this equals\footnote{At the non--separating node,
\be \Gamma_{d,d;2} (\tau_1,\tau_2,\tau) \rightarrow ({\rm Im}\tau_2)^{d/2}\Gamma_{d,d;1}(\tau_1).\ee}
\be 2\int d^2\tau_1 d^2\tau \int_{-1/2}^{1/2} d{\rm Re} \tau_2 ({\rm Im}\tau_1)^{-3} ({\rm Im}\tau_2)^{-1}\frac{\p}{\p{\rm Im}\tau_2} ({\rm Im}\tau_2)^{d/2} \Gamma_{d,d;1}(\tau_1)\Big\vert_{{\rm Im}\tau_2 = L\rightarrow \infty},\ee
where we have also divided by a factor of two because of the $\mathbb{Z}_2$ automorphism of the remaining $T^2$ with complex structure $\tau_1$ (this will also be implicit in the relevant cases below). This yields a finite contribution only when $d=4$. Finally, we perform the $\tau$ integral easily using
\be -\frac{1}{2} \leq {\rm Re}\tau \leq \frac{1}{2}, \quad -\frac{{\rm Im}\tau_1}{2} \leq {\rm Im}\tau \leq \frac{{\rm Im}\tau_1}{2}.\ee
Thus $\tau$ is a point on the worldsheet of the $T^2$ with complex structure $\tau_1$. This leads to
\be \label{modinv}\Big(\Delta_{O(d,d,\mathbb{Z})} +d(d-3)\Big) \hat{I}^{(2)}_{D^4\mathcal{R}^4} = 8 \delta_{d,4}\int_{\mathcal{F}_1} \frac{d^2\tau_1}{({\rm Im}\tau_1)^2} \Gamma_{d,d;1},\ee
and hence
\be \frac{\pi}{2}\Big(\Delta_{O(d,d,\mathbb{Z})} +d(d-3)\Big) \hat{I}^{(2)}_{D^4\mathcal{R}^4} =4 I^{(1)}_{\mathcal{R}^4} \delta_{d,4}.\ee
This contribution entirely comes from the first term involving $\p/\p\tau_2$ in the expression for $\mathcal{D}_{\tau_2}$ in \C{defD}. While the third term involving $\p/\p\tau$ trivially vanishes, the second term involving $\p/\p\tau_1$ yields a divergent contribution from the boundary of $\tau_1$, given by
\be \label{none}\frac{1}{3}\delta_{d,6} \int_{\mathcal{F}_1} d^2\tau_1 \frac{\p \Gamma_{6,6;1}}{\p {\rm Im}\tau_1} = \frac{2}{3}\delta_{d,6}\int_{-1/2}^{1/2} d{\rm Re}\tau_1 \Gamma_{6,6;1}\Big\vert_{{\rm Im}\tau_1 =L\rightarrow \infty} = \frac{2L^3}{3}\delta_{d,6},\ee
and hence does not contribute to the final equation. These kind of contributions which vanish in the final answer, will be ignored in general.

As discussed before, there is another contribution that we need to include to complete the analysis. This is the field theory limit where both ${\rm Im}\tau_1 \rightarrow \infty$ and ${\rm Im}\tau_2 \rightarrow \infty$. As above in the expression for $\mathcal{D}_{\tau_2}$ we choose the term involving $\p/\p\tau_1$ to obtain       
\be \frac{1}{3 ({\rm Im}\tau_2)^{3-d/2}} \int d^2 \tau_1 \frac{\p}{\p{\rm Im}\tau_1}\Gamma_{d,d;1} (\tau_1) \Big\vert_{{\rm Im}\tau_2 \rightarrow \infty} \ee
where we have integrated over $\tau$. This receives non--vanishing contribution only from the boundary at ${\rm Im}\tau_1\rightarrow\infty$ which is the field theory limit, to give us
\be  \frac{2({\rm Im}\tau_1)^{d/2}}{3 ({\rm Im}\tau_2)^{3-d/2}}\Big\vert_{{\rm Im}\tau_i \rightarrow \infty}.\ee  
For the special case when $d=3$, this reduces to
\be \frac{2}{3} \Big( \frac{{\rm Im}\tau_1}{{\rm Im}\tau_2}\Big)^{3/2}\Big\vert_{{\rm Im}\tau_i \rightarrow \infty} = {\rm constant},\ee
since ${\rm Im}\tau_1/{\rm Im}\tau_2 \sim O(1)$. Hence we have that
\be \frac{\pi}{2}\Big(\Delta_{O(d,d,\mathbb{Z})} +d(d-3)\Big) \hat{I}^{(2)}_{D^4\mathcal{R}^4} =4 I^{(1)}_{\mathcal{R}^4} \delta_{d,4} +c\delta_{d,3}\ee
where $c$ is a constant. Rather than doing a string computation to determine $c$, we can determine it in the field theory limit much like the analysis of the genus one $\mathcal{R}^4$ amplitude and genus two $D^6\mathcal{R}^4$ amplitude in eight dimensions. This is because the seven dimensional effective action has a non--analytic term of the schematic form~\cite{Bern:1998ug,Green:2010sp}
\be l_s^5 \int d^7 x \sqrt{-g^{(7)}} (e^{-2\phi} V_3)^{-1}{\rm ln}(-\mu \alpha' s) D^4\mathcal{R}^4\ee
in the string frame at genus two. This is obtained from the logarithmic ultraviolet divergence of two loop supergravity. On converting to the Einstein frame, it yields a dilaton dependent local contribution given by
\be \label{pi2}\frac{16\pi^2}{15} {\rm ln} g_3\ee   
to the $D^4\mathcal{R}^4$ interaction. This is a part of the perturbative expansion of the $SL(5,\mathbb{Z})$ invariant $D^4 \mathcal{R}^4$ coefficient function $\mathcal{E}_{D^4\mathcal{R}^4}$ which satisfies
\be (\Delta_\mathcal{U} +\ldots )\mathcal{E}_{D^4 \mathcal{R}^4} = 40\zeta(2),\ee 
where $\Delta_\mathcal{U}$ is the $SL(5,\mathbb{Z})$ invariant Laplacian\footnote{The eigenvalue vanishes for $d=3$, hence the answer is independent of the scale of the logarithm in \C{pi2}.}. Now from \C{arbd} we get that
\be \Delta_\mathcal{U} = \frac{5}{8} \p^2_{\phi_3}+\frac{5}{2} \p_{\phi_3}+\Delta_{O(3,3,\mathbb{Z})} +\ldots,\ee
leading to 
\be c=24\zeta(2).\ee

We now consider the genus two $D^6\mathcal{R}^4$ amplitude.  
Proceeding along lines similar to the earlier analysis, we see that the $D^6\mathcal{R}^4$ amplitude satisfies
\be \label{Arb2}\Big(\Delta_{O(d,d,\mathbb{Z})} +(d+2)(d-5)\Big) \hat{I}^{(2)}_{D^6\mathcal{R}^4} =-\frac{(I^{(1)}_{\mathcal{R}_4})^2}{\pi}+\ldots,\ee
where we have considered only the contribution coming from the separating node. The contribution to the right hand side of \C{Arb2} from the non--separating node in figure 1 is given by
\be \label{van}2i \int d^2\tau_1 d^2\tau \Big[ \varphi \mathcal{D}_{\tau_2}({\rm Im}\tau_2)^{d/2}\Gamma_{d,d;1}(\tau_1)- ({\rm Im}\tau_2)^{d/2}\Gamma_{d,d;1}(\tau_1)\mathcal{D}_{\tau_2} \varphi - c.c. \Big]\Big\vert_{{\rm Im}\tau_2 = L\rightarrow \infty}.\ee 
First we consider the $O({\rm Im}\tau_2)$ term in the expression for the KZ invariant in \C{nonsep}. Using the first term in the expression for $\mathcal{D}_{\tau_2}$ in \C{defD} involving $\p/\p\tau_2$, this can potentially lead to a non--vanishing contribution only for $d=2$. However, the two contributions in \C{van} exactly cancel\footnote{The contributions are of the form $\delta_{d,2} I^{(1)}_{\mathcal{R}^4}$.}. For the term in \C{defD} involving $\p/\p\tau_1$ the analysis proceeds exactly as in the $D^4\mathcal{R}^4$ case above leading to a term of the form \C{none} with $\delta_{d,4} \Gamma_{4,4;1}$ instead. Again, its contribution vanishes. Finally, one obtains a non--vanishing contribution in the field theory limit in eight dimensions as has been discussed before.   

Next we consider the $O(1)$ term in \C{nonsep}. Using the $\p/\p\tau_2$ term in the expression for $\mathcal{D}_{\tau_2}$ on the right hand side of \C{Arb2}, this contribution from the non--separating node equals
\be \label{invmod}4\delta_{d,4} \int_{\mathcal{F}_1} \frac{d^2\tau_1}{({\rm Im}\tau_1)^3} \Gamma_{d,d;1}(\tau_1)\mathcal{I}(\tau_1),\ee
where
\be \label{mI}\mathcal{I} (\tau_1) = \int d^2\tau \Big[ \frac{5\pi({\rm Im}\tau)^2}{6({\rm Im}\tau_1)}-{\rm ln} \Big\vert \frac{\theta_1 (\tau,\tau_1)}{\eta(\tau_1)}\Big\vert \Big] .\ee
Thus \C{mI} is an integral over the worldsheet of a torus with complex structure $\tau_1$, where $\tau$ is the worldsheet coordinate. 
Now
\bea &&\frac{5\pi({\rm Im}\tau)^2}{6({\rm Im}\tau_1)}-{\rm ln} \Big\vert \frac{\theta_1 (\tau,\tau_1)}{\eta(\tau_1)}\Big\vert \non \\ &&= 2 P(\tau,\tau_1) -\frac{\pi({\rm Im}\tau)^2}{6({\rm Im}\tau_1)} -{\rm ln} \vert \sqrt{2\pi} \eta(\tau_1)\vert^2,\eea
where $P(\tau,\tau_1)$ is the scalar propagator on the torus  with complex structure $\tau_1$, where the points are separated by $\tau$, given by
\be P(\tau,\tau_1) = -\frac{1}{4} {\rm ln} \Big\vert \frac{\theta_1 (\tau,\tau_1)}{\theta_1'(\tau_1)}\Big\vert^2 +\frac{\pi({\rm Im}\tau)^2}{2({\rm Im}\tau_1)},\ee
where we have used $\theta'_1(\tau_1) = -2\pi\eta(\tau_1)^3$.
Now~\cite{Green:1999pv}\footnote{$\hat{P}$ is proportional to the Kronecker--Eisenstein series $D_{1,1}$.}
\bea &&2\hat{P} (\tau,\tau_1) \equiv 2 P(\tau,\tau_1)  -{\rm ln} \vert \sqrt{2\pi} \eta(\tau_1)\vert^2 \non \\ &&= \frac{1}{2\pi}\sum_{(m,n)\neq (0,0)} \frac{{\rm Im} \tau_1}{\vert m\tau_1 +n\vert^2} e^{\pi[\bar\tau(m\tau_1 +n)- \tau(m\bar\tau_1 +n)]/{\rm Im} \tau_1},\eea
which has the useful property that it vanishes when integrated over the toroidal worldsheet\footnote{In fact, 
\be \hat{P} (z,\tau) = -\frac{1}{2} {\rm ln}g(z,\tau),\ee
where ${\rm ln}g(z,\tau)$ is the genus one Arakelov Green function.}  
\be \int d^2\tau \hat{P} (\tau,\tau_1)=0.\ee
Thus
\be \mathcal{I} (\tau_1) = -\frac{\pi({\rm Im}\tau_1)^2}{72}.\ee
This does not lead to a modular invariant contribution to \C{invmod} and hence its contribution vanishes in the final answer. Similar is the conclusion for other contributions that result at this order, and we get no new type of contribution.

We next consider the power suppressed term of order $({\rm Im}\tau_2)^{-1}$ in the expression in \C{nonsep}. Proceeding as above, we find a modular invariant contribution from the $\p/\p\tau_2$ part of $\mathcal{D}_{\tau_2}$ equal to
\be \label{calc}8\delta_{d,6} \int_{\mathcal{F}_1} \frac{d^2\tau_1}{({\rm Im}\tau_1)^3} \Gamma_{d,d;1} (\tau_1)\int d^2\tau \varphi_1 (\tau_1,\tau).\ee
On using
\be \int d^2 \tau D_{2,2} (\tau_1,\tau) =0, \ee
we see that \C{calc} equals
\be \frac{20}{\pi^3} \delta_{d,6} \int_{\mathcal{F}_1} \frac{d^2\tau_1}{({\rm Im}\tau_1)^2} E_2 (\tau_1,\bar\tau_1)\Gamma_{d,d;1} = \frac{20}{\pi^2} \delta_{d,6} I^{(1)}_{D^4\mathcal{R}^4}.\ee
There is no other contribution at this order.

We now consider the term at order $({\rm Im}\tau_2)^{-2}$ in the expression in \C{nonsep} given by
\be \frac{({\rm Im}\tau)^2 \varphi_1}{({\rm Im}\tau_2)^2 {\rm Im}\tau_1}.\ee
Though this contribution is considerably suppressed, it yields a non--vanishing contribution to \C{van} when the term involving $\p/\p\tau_1$ in the expression for $\mathcal{D}_{\tau_1}$ is considered. This arises entirely from the $\pi\zeta(3)/{\rm Im}\tau_1$ term in the expression for $E_2 (\tau_1,\bar\tau_1)$ in $\varphi_1$. This contribution arises from the first term in \C{van} and is given by\be \frac{\zeta(3)}{16\pi^2({\rm Im}\tau_2)^{5-d/2}} \int_{\mathcal{F}_1} d^2\tau_1 \frac{\p\Gamma_{d,d;1}(\tau_1)}{\p{\rm Im}\tau_1}\Big\vert_{{\rm Im}\tau_2\rightarrow \infty}.\ee 
This receives a non--vanishing contribution only as ${\rm Im}\tau_1 \rightarrow \infty$ for $d=5$ and corresponds to the field theory limit, with a contribution given by
\be \frac{\zeta(3)}{8\pi^2} \delta_{d,5}\Big(\frac{{\rm Im}\tau_1}{{\rm Im}\tau_2}\Big)^{5/2}\Big\vert_{{\rm Im}\tau_i\rightarrow \infty}.\ee 
Note that an overall factor of $\zeta(3)$ automatically arises in the analysis. 

Thus we get that
\be \pi\Big(\Delta_{O(d,d,\mathbb{Z})} +(d+2)(d-5)\Big) \hat{I}^{(2)}_{D^6\mathcal{R}^4} =-(I^{(1)}_{\mathcal{R}_4})^2+\frac{20}{\pi} \delta_{d,6} I^{(1)}_{D^4\mathcal{R}^4}+ \zeta(3) d \delta_{d,5}-\frac{13\zeta(2)}{6}\delta_{d,2},\ee
where $d$ is a constant that should be fixed by an analysis of two loop supergravity. However, this analysis has not been done, and we simply use the expressions based on U--duality~\cite{Pioline:2015yea} to fix $d$.  

Based on U--duality, on converting to the Einstein frame, we get a dilaton dependent local contribution given by
\be  \frac{40\zeta(3)}{9g_5^4} {\rm ln} g_5\ee   
to the $D^6\mathcal{R}^4$ interaction. This is a part of the perturbative expansion of the $E_{6(6)}$ invariant $D^6 \mathcal{R}^4$ coefficient function $\mathcal{E}_{D^6\mathcal{R}^4}$ which satisfies~\cite{Pioline:2015yea}\footnote{The $O(g_5^{-2})$ term will be useful later on.}
\be \label{nl}(\Delta_\mathcal{U} +\ldots )\mathcal{E}_{D^6 \mathcal{R}^4} = \frac{110\zeta(3)}{3g_5^4} +\frac{55 I^{(1)}_{\mathcal{R}^4}}{3g_5^2}+\ldots,\ee 
where $\Delta_\mathcal{U}$ is the $E_{6(6)}$ invariant Laplacian. Now from \C{arbd} we get that
\be \Delta_\mathcal{U} = \frac{3}{8} \p^2_{\phi_5}+6 \p_{\phi_5}+\Delta_{E_{6(6)}} +\ldots,\ee
leading to 
\be d=\frac{70}{3}.\ee 
There are no other contributions at this order.
There are neither as one considers higher order terms in $({\rm Im}\tau_2)^{-1}$ in \C{nonsep} as well.

\subsection{The genus three amplitude}

At genus three, the $D^6\mathcal{R}^4$ interaction is given by \C{G3}. Making use of the relation~\cite{Obers:1999um}
\be \Big(\Delta_{O(d,d,\mathbb{Z})} +\frac{3d(d-4)}{2}\Big) \Gamma_{d,d;3} = 2\Delta_{Sp(6,\mathbb{Z})} \Gamma_{d,d;3},\ee
we get that
\be\label{rHs}\Big(\Delta_{O(d,d,\mathbb{Z})} +\frac{3d(d-4)}{2}\Big)\hat{I}^{(3)}_{D^6\mathcal{R}^4} =2\int_{\mathcal{F}_3} d\mu_3 \Delta_{Sp(6,\mathbb{Z})}\Gamma_{d,d;3},\ee
yielding a boundary term on the right hand side. We denote the period matrix by
\be \Omega= \left( \begin{array}{ccc} \tau_1 & \tau &\s \\ \tau & \tau_2 &\rho \\ \s&\rho &\tau_3\end{array} \right).\ee
Thus the $1,2$-th block of the matrix is given by the genus two period matrix \C{parap} which we refer to as $\Omega_2$. Importantly, from \C{defdom}, we see that
\be \label{C}2\vert {\rm Im}\tau\vert \leq {\rm Im}\tau_1 , \quad 2\vert {\rm Im}\s\vert \leq {\rm Im}\tau_1 , \quad 2\vert {\rm Im}\rho\vert \leq {\rm Im}\tau_2, \quad {\rm Im} \tau_1 \leq {\rm Im}\tau_2 \leq {\rm Im}\tau_3 .\ee

\begin{figure}[ht]
\begin{center}
\[
\mbox{\begin{picture}(150,70)(0,0)
\includegraphics[scale=.45]{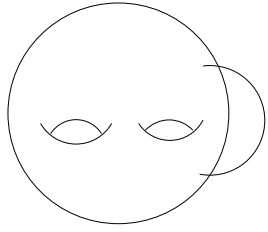}
\end{picture}}
\]
\caption{The non--separating genus three node}
\end{center}
\end{figure}

We now consider the boundary terms that arise on the right hand side of \C{rHs} where we use the relation
\be ({\rm det} Y)^{-4}\Delta_{Sp(6,\mathbb{Z})} = 2\bar\p_{IJ} \Big(({\rm det}Y)^{-4} Y_{IK} Y_{JL} \p_{KL}\Big) +c.c. .\ee

As discussed before, there are no contributions from the separating nodes due to the lack of a pole term in the integrand. Hence the only non--vanishing contribution arises from the boundary of moduli space as ${\rm Im}\tau_3\rightarrow \infty$, which is a non--separating node as shown in figure 2. This contributes
\be \label{loc1}4 \int_{\mathcal{F}_3} d^2 \tau_1 d^2\tau_2 d^2\tau_3 d^2\s d^2\rho d^2\tau \bar\p_{\tau_3} \Big(({\rm det}Y_2)^{-4} ({\rm Im}\tau_3)^{-4} Y_{3I} Y_{3J} \p_{IJ}\Big)({\rm Im}\tau_3)^{d/2}\Gamma_{d,d;2}(\Omega_2) +c.c. ,\ee
where we have replaced ${\rm det}Y$ by $({\rm det}Y_2)(\rm Im\tau_3)$ as this is the leading contribution as ${\rm Im}\tau_3 \rightarrow \infty$ on which the integral in \C{loc1} localizes. We have also used
\be \Gamma_{d,d;3}(\Omega) \rightarrow ({\rm Im}\tau_3)^{d/2} \Gamma_{d,d;2}(\Omega_2)\ee
in this limit, which is independent of $\s$ and $\rho$. We first consider the contribution coming from $\p_{IJ} \sim \p_{\tau_3}$ in \C{loc1}. This equals\footnote{We have divided by a factor of two since every genus two surface has a $\mathbb{Z}_2$ symmetry.}  
\be \label{d6}\frac{8}{({\rm Im}\tau_3)^2} \frac{\p({\rm Im}\tau_3)^{d/2}}{\p{\rm Im}\tau_3} \int_{\mathcal{F}_2} \frac{d^2 \tau_1 d^2\tau_2  d^2\tau}{({\rm det}Y_2)^3}\Gamma_{d,d;2}(\Omega_2) \ee
as ${\rm Im}\tau_3 \rightarrow \infty$ where we have integrated over $\s$ and $\rho$. We have also set $({\rm Im}\tau_1)({\rm Im}\tau_2)/{\rm det}Y_2 =1$ as the corrections involving powers of $({\rm Im}\tau)^2/({\rm Im}\tau_1)({\rm Im}\tau_2)$ do not yield $Sp(4,\mathbb{Z})$ invariant integrands and hence must vanish in the final answer. Thus \C{d6} is non--vanishing only in four dimensions, leading to
\be 24 \delta_{d,6} \int_{\mathcal{F}_2} d\mu_2 \Gamma_{d,d;2}(\Omega_2) = \frac{48}{\pi} \delta_{d,6} I^{(2)}_{D^4\mathcal{R}^4}.\ee

What about the other contributions in \C{loc1}? They are localized at ${\rm Im}\tau_3 \rightarrow \infty$, and can only yield a non--vanishing contribution from the boundary of moduli space as ${\rm Im}\tau_2 \rightarrow \infty$, and so we consider the term where $\p_{IJ} \sim \p_{\tau_2}$. In this analysis and the one that follows, we shall drop all numerical prefactors. We get a contribution  
\bea &&\frac{1}{({\rm Im}\tau_3)^{4-d/2}} \int d^2 \tau_1 d^2\tau_2 d^2\s d^2\rho d^2\tau \frac{({\rm Im}\rho)^2}{({\rm det}Y_2)^4} \p_{\tau_2} \Gamma_{d,d;2}(\Omega_2)\Big\vert_{{\rm Im}\tau_3\rightarrow \infty} +c.c. \non \\ &&\sim \frac{1}{({\rm Im}\tau_3)^{4-d/2}} \int \frac{d^2 \tau_1 d^2\tau_2  d^2\tau}{({\rm Im}\tau_1)^3} \p_{\tau_2} \Big(\frac{\Gamma_{d,d;2}(\Omega_2)}{{\rm Im}\tau_2}\Big)\Big\vert_{{\rm Im}\tau_3\rightarrow \infty}+c.c.  .\eea
We have again replaced ${\rm det Y_2}$ by $({\rm Im}\tau_1)({\rm Im}\tau_2)$, and moved $({\rm Im}\tau_2)^{-1}$ into the total derivative as the remaining terms do not yield a finite contribution. Thus the integral receives contributions only from the boundary of moduli space as ${\rm Im}\tau_2 \rightarrow \infty$ and yields  
\be \label{ft}\frac{({\rm Im}\tau_2)^{d/2-1}}{({\rm Im}\tau_3)^{4-d/2}}\Big\vert_{{\rm Im}\tau_i\rightarrow \infty} \int_{\mathcal{F}_1} \frac{d^2\tau_1}{({\rm Im}\tau_1)^2} \Gamma_{d,d;1}(\tau_1),\ee
which yields a finite contribution only in five dimensions. This leads to a contribution of the form
\be \frac{16e_1}{5} \delta_{d,5} I^{(1)}_{\mathcal{R}^4},\ee
as depicted in figure 1, where $e_1$ is a constant.

Finally, we further consider the contribution from the boundary of moduli space of $\tau_1$ when the resulting answer yields the field theory limit. From \C{ft}, we see this contribution is of the form
\be \frac{({\rm Im}\tau_2)^{d/2-1}({\rm Im}\tau_2)^{d/2-1}}{({\rm Im}\tau_3)^{4-d/2}}\Big\vert_{{\rm Im}\tau_i\rightarrow \infty} \ee
which produces a finite contribution in six dimensions, to yield the result
\be \frac{16e_2}{5} \delta_{d,4},\ee
where $e_2$ is a constant. Note that this argument does not determine exactly how the various limits are taken, though it yields the correct count. One has to take the limits such that the Mercedes skeleton diagram of three loop supergravity is obtained, not the ladder skeleton diagram which has the $D^8\mathcal{R}^4$ term as the leading term in the momentum expansion~\cite{Bern:2008pv}\footnote{We have not calculated directly from the expression for the string amplitude these types of contributions which involve ratios of limits of moduli each of which diverge, though the dimensions where they are nonvanishing is determined by our analysis. It would be interesting to fix these coefficients directly in string theory.}.

Thus we get that
\be \frac{5}{16}\Big(\Delta_{O(d,d,\mathbb{Z})} +\frac{3d(d-4)}{2}\Big)\hat{I}^{(3)}_{D^6\mathcal{R}^4} =e_2 \delta_{d,4} + e_1 \delta_{d,5} I^{(1)}_{\mathcal{R}^4} +\frac{15}{\pi} \delta_{d,6} I^{(2)}_{D^4\mathcal{R}^4}.\ee

We first fix $e_2$ using the fact that
the six dimensional effective action has a non--analytic term of the schematic form~\cite{Bern:2008pv,Green:2010sp}
\be l_s^8 \int d^6 x \sqrt{-g^{(6)}} (e^{-2\phi} V_3)^{-2}{\rm ln}(-\mu \alpha' s) D^6\mathcal{R}^4\ee
in the string frame at genus three. This is obtained from the logarithmic ultraviolet divergence of three loop supergravity. On converting to the Einstein frame, it yields a dilaton dependent local contribution given by
\be 5\zeta(3) {\rm ln} g_4\ee   
to the $D^6\mathcal{R}^4$ interaction. This is a part of the perturbative expansion of the $SO(5,5,\mathbb{Z})$ invariant $D^6 \mathcal{R}^4$ coefficient function $\mathcal{E}_{D^6\mathcal{R}^4}$ which satisfies
\be (\Delta_\mathcal{U} +\ldots )\mathcal{E}_{D^6 \mathcal{R}^4} = 40\zeta(3),\ee 
where $\Delta_\mathcal{U}$ is the $SO(5,5,\mathbb{Z})$ invariant Laplacian\footnote{Once again, the eigenvalue vanishes in six dimensions.}. Now
\be \Delta_\mathcal{U} = \frac{1}{2} \p^2_{\phi_4}+4 \p_{\phi_4}+\Delta_{O(4,4,\mathbb{Z})} +\ldots,\ee
leading to 
\be e_2 = 20\zeta(3).\ee

To fix $e_1$, in the absence of the supergravity analysis, we use the results obtained using U--duality~\cite{Pioline:2015yea}. On converting to the Einstein frame, we get a dilaton dependent local contribution given by
\be  \frac{20I^{(1)}_{\mathcal{R}^4}}{9g_5^2} {\rm ln} g_5\ee   
to the $D^6\mathcal{R}^4$ interaction. Then from \C{nl} we get that
\be e_1 = \frac{25}{3}.\ee
 In the various analysis that we have done for these BPS amplitudes in various dimensions, the results we have obtained agree with results based on U--duality~\cite{Basu:2007ru,Basu:2007ck,Green:2010wi,Green:2010kv,Pioline:2015yea}. 

Thus we obtain the complete Poisson equations satisfied by the $\mathcal{R}^4$, $D^4\mathcal{R}^4$ and $D^6\mathcal{R}^4$ interactions in toroidal compactifications of type II string theory that preserve maximal supersymmetry. Our results agree with several proposals for these amplitudes in the literature, and hence prove certain results. The general strategy is to obtain Poisson equations satisfied by these amplitudes, where the source terms are entirely determined by contributions from the boundary of moduli space, which involve only the asymptotic behavior of the various integrands. While we saw that there are many complicated looking terms, most of them do not produce finite contributions of the desired kind. They either diverge or do not lead to modular invariant expressions integrated over moduli space which can be interpreted as contributions from string amplitudes at lower orders in the genus expansion. Hence such contributions must cancel in the final answer, from terms involving expanding around various nodes. Thus modular invariance of the string amplitudes plays an important role in ruling out various contributions. Thus at the end the source terms involve contributions from amplitudes at lower orders in the genus expansion which involve an integration over a part of the original moduli space in a modular invariant way, or moduli independent contributions involving only the massless modes whose structure is determined by perturbative supergravity. The final equations take a simple form as the amplitudes belong to BPS couplings in the low energy effective action.

It would be interesting to generalize the analysis to string amplitudes in theories with less supersymmetry, at least in the BPS sector. One could also use these techniques to analyze the structure of the non--BPS amplitudes, like the $D^8\mathcal{R}^4$ interaction to start with. While higher genus expressions for such amplitudes are not known, multi--loop supergravity analysis might provide useful hints about their perturbative structure, along the lines of~\cite{Bern:2009kd,Basu:2014uba,Basu:2015dsa}.

\providecommand{\href}[2]{#2}\begingroup\raggedright\endgroup

\end{document}